\documentclass[%
twocolumn,
 amsmath,amssymb,
 aps,
 superscriptaddress,
pra,
]{revtex4-2}

\usepackage{graphicx}
\usepackage{dcolumn}
\usepackage{bm}
\usepackage{svg}
\usepackage{upgreek}
\usepackage[hidelinks]{hyperref}
\usepackage[capitalise]{cleveref}
\usepackage{amssymb}
\usepackage{float}
\usepackage[normalem]{ulem}
\usepackage[colorinlistoftodos]{todonotes}
 
\usepackage{braket}

\usepackage{color}
\usepackage{soul,xcolor}
\definecolor{DarkGreen}{rgb}{0, 0.5, 0}

\setstcolor{red}

\makeatletter
\newsavebox{\@brx}
\newcommand{\llangle}[1][]{\savebox{\@brx}{\(\m@th{#1\langle}\)}%
  \mathopen{\copy\@brx\kern-0.5\wd\@brx\usebox{\@brx}}}
\newcommand{\rrangle}[1][]{\savebox{\@brx}{\(\m@th{#1\rangle}\)}%
  \mathclose{\copy\@brx\kern-0.5\wd\@brx\usebox{\@brx}}}
\makeatother

\newboolean{showkeys}
\setboolean{showkeys}{false}
\ifthenelse{\boolean{showkeys}}{\usepackage[notref,notcite]{showkeys}} 
{}

\usepackage{scalerel}
\usepackage{stackengine,wasysym}

\newcommand{\ident}{{\mathbb{I}}}

\newcommand{\approptoinn}[2]{\mathrel{\vcenter{
  \offinterlineskip\halign{\hfil$##$\cr
    #1\propto\cr\noalign{\kern2pt}#1\sim\cr\noalign{\kern-2pt}}}}}

\begin{document}

\title{Identifying unbound strong bunching and the breakdown of the Rotating Wave Approximation in the quantum Rabi model}

 	\author{\'Alvaro Nodar}
     \affiliation{Materials Physics Center, CSIC-UPV/EHU, 20018 Donostia-San Sebasti\'{a}n, Spain}
     \affiliation{Donostia International Physics Center DIPC, 20018 Donostia-San Sebasti\'{a}n, Spain}
	\affiliation{School of Mathematical and Physical Sciences, Macquarie University, NSW 2109, Australia.}
	
     \author{Ruben Esteban}
     \email{ruben\_esteban@ehu.eus}
     \affiliation{Materials Physics Center, CSIC-UPV/EHU, 20018 Donostia-San Sebasti\'{a}n, Spain}
     \affiliation{Donostia International Physics Center DIPC, 20018 Donostia-San Sebasti\'{a}n, Spain}
    
     \author{Unai Muniain}
     \affiliation{Donostia International Physics Center DIPC, 20018 Donostia-San Sebasti\'{a}n, Spain}

    \author{Michael J. Steel}
 	\email{michael.steel@mq.edu.au}
	\affiliation{School of Mathematical and Physical Sciences, Macquarie University, NSW 2109, Australia.}

     \author{Javier Aizpurua}
     \email{aizpurua@ehu.eus}
     \affiliation{Materials Physics Center, CSIC-UPV/EHU, 20018 Donostia-San Sebasti\'{a}n, Spain}
     \affiliation{Donostia International Physics Center DIPC, 20018 Donostia-San Sebasti\'{a}n, Spain}

    \author{Miko\l{}aj K. Schmidt}
 	\email{mikolaj.schmidt@mq.edu.au}
	\affiliation{School of Mathematical and Physical Sciences, Macquarie University, NSW 2109, Australia.}

\date{\today}

\begin{abstract}
We use a recently derived gauge-invariant formulation of the problem of a two-level system coupled to an optical cavity, to explore the transition between the weak, and the ultra-strong coupling regimes of light-matter interaction. We explore this transition using the intensity correlations $g^{(2)}(\tau)$ of the emitted light, and find strong, unbounded bunching of the emission from systems governed by the Rabi Hamiltonian. Surprisingly, this effect is observed not only in the ultra-strong coupling regime, but also for weakly coupled systems, where the Jaynes-Cummings Hamiltonian would predict the opposite, antibunched emission. This suggests that the higher-order correlations are a particularly sensitive probe of the divergence between the Jaynes-Cummings and Rabi Hamiltonians, and can serve as an indicator of the breakdown of the rotating wave approximation. Our findings indicate also that the boundary between the weakly, strongly, and ultra-strongly coupled dynamics, is much richer than currently accepted.
\end{abstract}

\maketitle

\section{Introduction}
Many systems studied in cavity quantum optics are variations of its fundamental workhorse: a two-level system (TLS) coupled to an optical cavity~\cite{scully_zubairy_1997, haroche2013nobel}. The behavior of the cavity-TLS system (CTS) is dictated by the relationships between the coupling strength ($g$) and the characteristic resonant frequencies ($\omega_\sigma$ and $\omega_a$) and dissipation rates ($\gamma$ and $\kappa$) of the TLS and the cavity, respectively. In particular, if $g$ is smaller than the dominant losses of the system ($g\lesssim (\kappa + \gamma)/2$, defining the \textit{weak coupling} (WC) regime), any energy that enters the system is likely lost before the exchange of excitations between the cavity and the TLS can occur. On the other hand, in the \textit{strong coupling} (SC) regime for which $g$ exceeds the losses of the system ($g\gtrsim (\kappa + \gamma)/2$), the cavity and the TLS can coherently exchange excitations before the energy is dissipated, inducing a hybridization in the response of the CTS \cite{torma2014strong, thompson1992observation, rempe1987observation}.

Large coupling strengths can lead the system into the so-called \textit{ultra-strong coupling} (USC) regime, where $g$ becomes comparable to the resonant frequencies of the cavity and TLS \footnote{We note these conventional criteria for the WC, SC, and USC regimes are not exclusionary — in particular, systems with very lossy cavities can simultaneously be characterized as weak, and ultra-strongly coupled if $\kappa \gtrsim 2 g \gtrsim \omega_a/5$ \cite{de2017virtual}.}. The phenomenological limit for the onset of the USC is typically defined as $g \gtrsim 0.1 \omega_a$, or $g \gtrsim 0.1 \omega_\sigma$ (USC systems are commonly studied in a resonant configuration where $\omega_a = \omega_\sigma$). Importantly, systems operating in the USC regime exhibit new characteristics and non-trivial properties, such as the existence of a non-vacuum ground state~\cite{frisk2019ultrastrong,RevModPhys.91.025005}, which would be largely missed if attempting to describe them using the typically approximations applied in the WC or SC regimes --- most importantly, the Rotating Wave Approximation (RWA). 

\begin{figure}[t!]
    \centering
    \includegraphics[width=\linewidth]{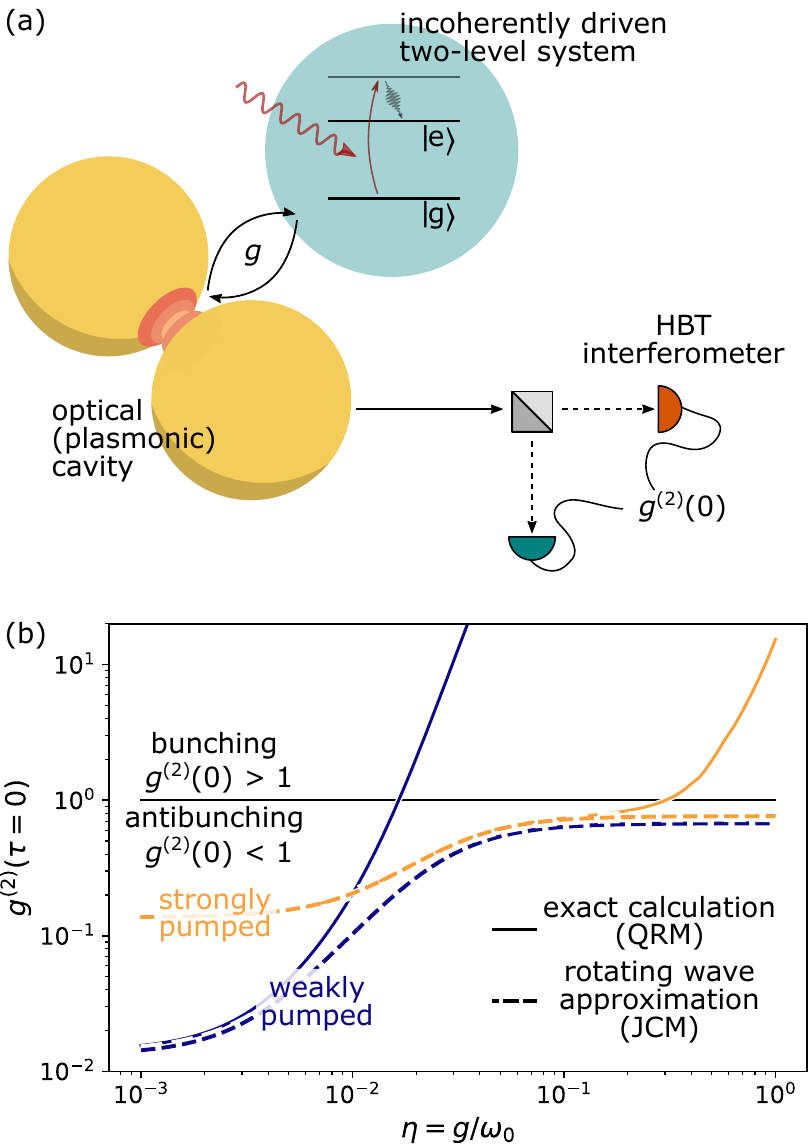}
    \caption{(a) Setup of the studied system. An incoherently driven TLS interacts with an optical cavity. The emission of the cavity is analyzed with a Hanbury Brown and Twiss (HBT) interferometer (further discussed in the text), giving the intensity correlations $g^{(2)}(\tau)$. (b) Intensity correlations $g^{(2)}(0)$ as a function of the coupling parameter $\eta=g/\omega_0$, calculated using the QRM (solid lines; see details in Section~\ref{subsec:QRM}) and the JCM (dashed lines; Section~\ref{subsec:JCM}). In the limit of strong incoherent pumping rate ($\Gamma/\gamma=10$, orange lines), the results of the two models coincide in the WC regime ($\eta \lesssim 2.5 \times 10^{-2}$), and become very different for USC $\eta\gtrsim 0.1$; for the weak pumping rate ($\Gamma/\gamma=10^{-6}$, dark blue lines), correlations obtained with the JCM and the QRM differ in the USC and WC regimes, for any $\eta \gtrsim 5 \times 10^{-3}$.}
    \label{fig:setup_problem}
\end{figure}

The emergence of new effects in the USC, and its boundary with WC and SC regimes, is typically quantified through the study of \textit{one-photon spectra} $S^{(1)}(\omega)$ of the emitted light, by tracing the intensities and frequencies of the spectral features, leading to the usual criterion for USC ($g\gtrsim 0.1 \omega_a$). Alternatively, we could choose to focus on the \textit{statistics}, rather than spectrum, of the emitted light, by measuring its intensity correlations $g^{(2)}(\tau)$, through the Hanbury Brown and Twiss (HBT) interferometer~\cite{hanbury_brown_apparent_1952,hanbury_brown_question_1956,kimble_photon_1977,carmichael1999statistical} shown schematically in Fig.~\ref{fig:setup_problem}(a). 
Intensity correlations should constitute a more sensitive probe to the non-number-conserving dynamics generated by the Rabi Hamiltonian~\cite{zubizarreta_casalengua_conventional_2020}, and might delineate an far more complex boundary of the USC regime. 

In this work, we analyze the emission of an incoherently-pumped CTS, operating in the WC, SC, and the USC regimes. We focus on studying the intensity correlations $g^{(2)}(\tau=0)$ obtained with two models: the Quantum Rabi model (QRM), and its widely used approximation --- the Jaynes-Cummings model (JCM). We embrace the formulation of the QRM recently derived in a series of papers which have reconciled  long-standing questions about ensuring gauge invariance~\cite{di_stefano2019resolution,Hughes2022gauge, settineri2018dissipation, PhysRevResearch.3.023079, babiker1983derivation,PhysRevResearch.4.023048}, and proposed a complete description of the interaction between an USC system and the environment. This formulation of the gauge-invariant QRM offers an opportunity to carefully study two questions, which are of critical importance to the field of cavity electrodynamics: (i) how does the statistics of emission from a CTS change as we transition between different coupling regimes? (ii) how does the JCM break down in the USC regime? 
To illustrate these effects, in Fig.~\ref{fig:setup_problem}(b) we plot $g^{(2)}(0)$ calculated using the QRM (solid lines) and the JCM (dashed lines), and two different pumping rates (details of the models and excitation schemes are discussed in Section~\ref{sec:system}). In contrast to the reported works on the emission from a thermally pumped CTS~\cite{ridolfo2013nonclassical,Chen_2022}, we find that under incoherent illumination of the TLS, the emission of the CTS appears to exhibit a seemingly unbounded bunching $g^{(2)}(0)\gg 1$~\footnote{The formal definitions of bunching and antibunching use the comparison between $g^{(2)}(0)$ and $g^{(2)}(\tau>0)$. Here we rely on the observation that for large delays $\tau$ to emission is necessarily uncorrelated ($g^{(2)}(\tau\rightarrow \infty)=1$) to identify bunching and antibunching with $g^{(2)}(0)>1$ and $g^{(2)}(0)<2$, respectively.}. Furthermore, we find that the two models (JCM and QRM) can deviate significantly both in the USC (see the dashed and solid orange lines diverging in Fig.~\ref{fig:setup_problem}(b) for $\eta = g/\omega_0 \gtrsim 0.1$), as well as in the WC regime (blue lines diverge for $\eta\sim 10^{-2}$; for the parameters used in Fig.~\ref{fig:setup_problem}, the WC is defined as $\eta \lesssim 2.5 \times 10^{-2}$, see Section~\ref{subsec:QRM} for details), depending on the incoherent pumping rate. 

The strong deviation between the JC and QR models, observed even in the WC regime in Fig.~\ref{fig:setup_problem}(b), is a surprising result with far-reaching consequences. The JCM is the default model for the majority of quantum-optical systems operating in the WC and SC regimes, including the circuit QED~\cite{RevModPhys.93.025005}, exciton polaritons~\cite{munoz2019emergence}, or quantum plasmonics~\cite{ZHOU20191}. It is therefore critical to understand how the Jaynes-Cummings model unravels when probed using higher-order correlations, by comparing it to more complete Quantum Rabi model, and carefully analyzing their relationship.

This work is structured as follows: in Section~\ref{sec:system} we formally introduce the system under study, the Quantum Rabi and Jaynes-Cumming Hamiltonians, and describe the formulation of the excitation, emission, and dissipation of the system. In Section~\ref{sec:origin.of.bunching} we identify the key mechanism which gives rise to this strong bunching. Finally, in Section~\ref{sec:extent} we probe the extent of this effect in the USC, SC, and WC regimes, and discuss how the intensity correlations can help us identify the breakdown of the RWA.

\section{Framework of the models}
\label{sec:system}

In this Section we briefly review the formulation of the Quantum Rabi (Subsection~\ref{subsec:QRM}) and Jaynes-Cummings (Subsection~\ref{subsec:JCM}) Hamiltonians. We describe how to formally address the interaction of such systems with the environment, and how to access the spectra, and the intensity correlations of the emitted light.

\subsection{Quantum Rabi model}
\label{subsec:QRM}

The exact description of the interaction between the cavity and the TLS is given by the Quantum Rabi Hamiltonian, which we choose to write in the Coulomb gauge~\cite{di_stefano2019resolution,
Hughes2022gauge}\footnote{We have verified that all results hold for the dipolar gauge.} as
\begin{align}\label{eq:HQRM}
    \hat{H} =~&\hbar \omega_a \hat{a}^\dagger \hat{a} \\ \nonumber
    &+\frac{\hbar\omega_\sigma}{2} \left\{ \hat{\sigma}_z \cos\left[2 \eta (\hat{a} + \hat{a}^\dagger)\right] + \hat{\sigma}_y \sin\left[ 2 \eta (\hat{a} + \hat{a}^\dagger)\right]\right\}.
\end{align}
{This form of the Hamiltonian was first derived by Di Stefano \textit{et al}. in Ref.~\citenum{di_stefano2019resolution}, and corrects the critical problem of the gauge-invariance required from the Hamiltonian.} 
Here $\omega_a$ and $\omega_\sigma$ are the resonant frequencies of the cavity and TLS, respectively. In this work, we consider a resonant system with $\omega_0\equiv \omega_a = \omega_\sigma$, and introduce the normalized coupling parameter $\eta = g / \omega_0$. Operator $\hat{a}$ denotes the cavity photon annihilation and $\hat{\sigma}_z = \hat{\sigma} \hat{\sigma}^\dagger - \hat{\sigma}^\dagger \hat{\sigma}$ and $\hat{\sigma}_y = i (\hat{\sigma}^\dagger - \hat{\sigma})$ are the $z$ and $y$ Pauli operators associated with the TLS. The $\hat{H}$ QR Hamiltonian does not conserve the number of \textit{excitations}, but conserves its \textit{parity}.

\begin{figure}[t!]
    \centering
    \includegraphics[width=.9\linewidth]{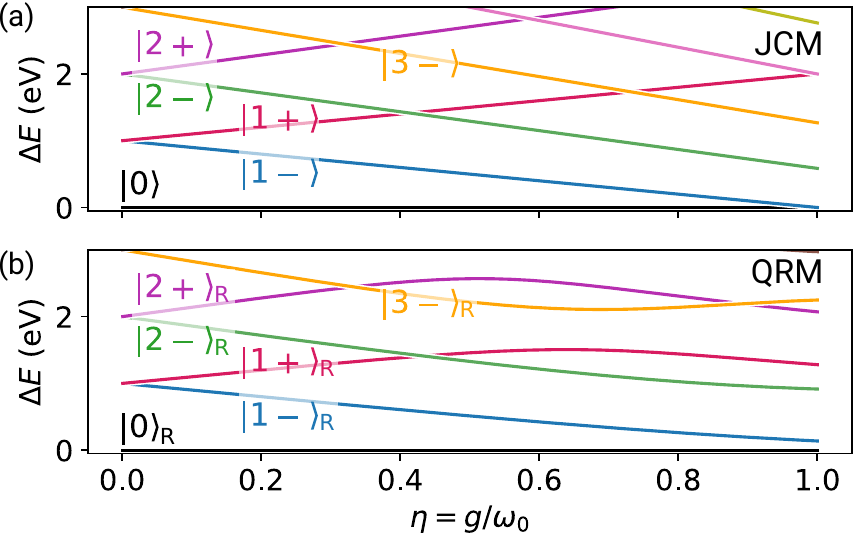}
    \caption{Eigenvalues obtained as function of the coupling strength within the (a) JCM and (b) QRM. For each $\ket{\nu}$ (for the JCM) or $\ket{\nu}_\text{R}$ (for the QRM) eigenstate, we show $\Delta E$: the difference between its eigenenergy ($\hbar \omega_\nu$) and the ground state energy ($\hbar \omega_0$).}
    \label{fig:Eigenvalues}
\end{figure}

To characterize the statistical properties of the emission of the CTS, we study the intensity correlations of the emitted photons, $g^{(2)}(\tau = 0)$, measured as
\begin{equation}
    g^{(2)}(\tau) = \frac{\braket{I_1(t+\tau) I_2(t)}}{\braket{I_1(t + \tau)}\braket{I_2(t)}},
    \label{eq:HBT}
\end{equation}
where $I_1(t+\tau)$ and $I_2(t)$ are the photocurrents registered by the two detectors of the HBT interferometer, and $\tau$ is the time delay between the detection events (see~Fig.~\ref{fig:setup_problem}(a)). 
For zero time delay ($\tau = 0$) and sufficiently large $t$ (so the system reaches the steady state), this quantity is related to the statistics of photons \textit{inside} the cavity as \cite{carmichael1999statistical}:
\begin{equation}
    g^{(2)}(0) = \frac{\braket{\hat{x}_a^\dagger\hat{x}_a^\dagger\hat{x}_a\hat{x}_a}_{ss}}{\braket{\hat{x}_a^\dagger \hat{x}_a}_{ss}^2},
    \label{eq:g2QRM}
\end{equation}
where $\braket{\hat{O}}_{ss}$ denotes the expectation value of operator $\hat{O}$ in the steady state ($ss$). The operator
\begin{equation}
    \hat{x}_a = \sum_{\nu, \mu; \omega_\nu > \omega_\mu} \ket{\mu}\!{}_\text{R}~{}_\text{R}\!\bra{\mu} i(\hat{a}^\dagger - \hat{a}) \ket{\nu}\!{}_\text{R}~{}_\text{R}\!\bra{\nu},
    \label{eq:dressed_a}
\end{equation}
is the \textit{dressed operator} of the cavity, which mediates the losses, and emission from the cavity~\cite{Hughes2022gauge, PhysRevResearch.3.023079}. 
{It is expressed in the basis of eigenstates $\ket{\mu}_\text{R}$ because the emission from the cavity occurs due to the transitions between the eigenstates of the Hamiltonian (see Refs. \citenum{settineri2018dissipation} and \citenum{PhysRevA.84.043832} for an in-depth discussion of this formulation and the role of the secular approximation).}
Kets $\ket{\nu}_\text{R}$ and $\ket{\mu}_\text{R}$ in Eq.~\eqref{eq:dressed_a} are the eigenvectors of the QRM Hamiltonian, and $\hbar \omega_\nu > \hbar \omega_\mu$ are their respective eigenvalues, plotted as a function of $\eta$ in Fig.~\ref{fig:Eigenvalues}(b). The notation for QRM eigenstates used throughout the text, $\ket{\nu}_\text{R} \equiv \ket{n\pm}_\text{R}$, is chosen to recall the JCM polaritons ($\ket{n\pm} = ( \ket{n, g} \pm \ket{n-1, e})/\sqrt{2}$), as the two match in the limit of vanishing coupling $g$ (see Fig.~\ref{fig:Eigenvalues}(a) and (b)). Note that throughout the text we keep the labeling of the eigenstates even after the eigenvalues crossing points (see color code in the figure).

The dynamics of the system, described through the density operator $\hat{\rho}$, is given by the master equation {with the gauge-invariant formulation of the Hamiltonian \cite{di_stefano2019resolution,PhysRevResearch.3.023079} and dissipation terms \cite{settineri2018dissipation,Hughes2022gauge,PhysRevA.84.043832}}:
\begin{align}
    \label{eq:MEQRM}
    \frac{\partial}{\partial t} {\hat{\rho}} &=  \mathcal{L} \hat{\rho}  \\ \nonumber
    &= \frac{1}{i\hbar}  [\hat{H}, {\hat{\rho}}]+\frac{\Gamma}{2} \mathcal{D}_{{\hat{x}_\sigma}^\dagger}[{\hat{\rho}}]  + \frac{\gamma}{2} \mathcal{D}_{{\hat{x}_\sigma}}[{\hat{\rho}}] + \frac{\kappa}{2} \mathcal{D}_{{\hat{x}_a}}[{\hat{\rho}}],
\end{align}
with $\mathcal{L}$ denoting the Lindbladian superoperator, and $\mathcal{D}_{\hat{O}}[\hat{\rho}] = (2 \hat{O}\hat{\rho}\hat{O}^\dagger  - \hat{O}^\dagger \hat{O} \hat{\rho}- \hat{\rho}\hat{O}^\dagger \hat{O})$ denoting the Gorini–Kossakowski–Sudarshan–Lindblad (GKSL) terms expressed through the dressed operators $\hat{x}_a$ (Eq.~\ref{eq:dressed_a}) and $\hat{x}_\sigma$ \cite{Hughes2022gauge}:
\begin{equation}
    \hat{x}_\sigma = \sum_{\nu, \mu; \omega_\nu > \omega_\mu} \ket{\mu}\!{}_\text{R}~{}_\text{R}\!\bra{\mu} (\hat{\sigma}^\dagger + \hat{\sigma}) \ket{\nu}\!{}_\text{R}~{}_\text{R}\!\bra{\nu}.
    \label{eq:dressed_s}
\end{equation}
The GKSL terms are introduced to describe the interaction between the TLS, cavity, and the bath following~\cite{PhysRevResearch.4.023048,Hughes2022gauge,PhysRevA.98.063828}, including modelling the incoherent pumping of the TLS (with rate $\Gamma$), the dissipation of the TLS ($\gamma$) and the cavity ($\kappa$). The dissipation is assumed additive, and we neglect any bath-mediated TLS-cavity interaction~\cite{settineri2018dissipation,frisk2019ultrastrong}. We define the steady state of the system via the density operator $\hat{\rho}_{ss}$ as $\partial_t \hat{\rho}_{ss}=0$. 

We choose the pumping mechanism and parameters which correspond to recent experiments with plasmonic CTS systems, which have reportedly reached values of $\eta$ close to 0.1~\cite{kuisma2022ultrastrong, usc_kenacohen, chikkaraddy_single-molecule_2016,pelton2019strong,melnikau2016rabi,bitton2022plasmonic, gambino2014exploring,Wei:13, gross2018near, gupta2021complex}. In particular, we explore the emission from a CTS under incoherent pumping of the TLS, corresponding to driving the atomic systems with a far blue-detuned laser, followed by a spontaneous cascade to the excited state of the TLS; see Fig.~\ref{fig:setup_problem}(a). In the plasmonic CTS considered here, the cavity is characterized by resonant frequency $\omega_0$ (set to $\hbar\omega_0=1$~eV), and a low quality factor $Q=\omega_0/\kappa$, defined by the dissipation rate $\kappa$, and set to $Q=20$. The emitter (such as a single molecule~\cite{kuisma2022ultrastrong, chikkaraddy_single-molecule_2016}, or a quantum dot \cite{gross2018near, bitton2022plasmonic, gupta2021complex}), is modelled as a TLS with decay rate $\gamma/\omega_0=10^{-3}$. These parameters establish the upper limit for the WC regime $\eta = g/\omega_0<\kappa/(2\omega_0) = 0.025$. For reference, we note that the characteristic cooperativities $C= 4 g^2 / (\kappa \gamma)$ in our study are $C = 0.08$ (for $\eta = 10^{-3}$) and $C = 800$ (for $\eta = 0.1$).

This framework is used to calculate the dependence of $g^{(2)}(0)$ (Eq.~\eqref{eq:g2QRM}) on the coupling parameter $\eta=g/\omega_0$, plotted in Fig.~\ref{fig:setup_problem}(b) as solid orange and blue lines, for the case of strong ($\Gamma/\gamma=10$), and weak ($\Gamma/\gamma=10^{-6}$) pumping, respectively. All calculations in this paper have been carried out using the Python package QuTiP~\cite{qutip1, qutip2}\footnote{We have considered in all of our QuTiP calculations an expansion of the Fock states of the cavity up to $N_a = 10$, which we verified ensures convergence.}.

\subsection{Jaynes-Cummings model}
\label{subsec:JCM}

The Jaynes-Cummings model (JCM) can be derived from the QRM by taking three approximations. First, we expand the interaction term in the QRM Hamiltonian as
\begin{align}
    &\hat{\sigma}_z \cos\left[2 \eta (\hat{a} + \hat{a}^\dagger)\right] + \hat{\sigma}_y \sin\left[ 2 \eta (\hat{a} + \hat{a}^\dagger)\right] \nonumber \\
    &= \hat{\sigma}_z + 2\eta \hat{\sigma}_y (\hat{a} + \hat{a}^\dagger) + O(\eta^2),
\end{align}
and drop the terms nonlinear in $\eta$. Next, we introduce the rotating wave approximation (RWA) by removing the so-called non-number-conserving terms $\hat{\sigma}\hat{a} + \hat{\sigma}^\dagger\hat{a}^\dagger$, to find the JCM Hamiltonian
\begin{equation}
    \hat{H}_\text{JC} = \hbar \omega_0 \hat{a}^\dagger \hat{a} + \hbar \frac{\omega_0}{2} \hat{\sigma}_z + i \hbar \omega_0 \eta (\hat{\sigma}^\dagger \hat{a} - \hat{a}^\dagger \hat{\sigma}).
    \label{eq:HJC}
\end{equation}
Figure~\ref{fig:Eigenvalues}(a) shows the eigenvalues of the JCM Hamiltonian which, for small $\eta$, converge with the eigenvalues of the QRM Hamiltonian (Fig.~\ref{fig:Eigenvalues}(b)).

The third approximation in the JCM regards the emission, dissipation, and absorption of the system, which, within the JCM, are mediated by the bare $\hat{a}$ and $\hat{\sigma}$ operators (instead of the dressed $\hat{x}_a$ and $\hat{x}_\sigma$ operators). The correlations arising in this model can thus be calculated as
\begin{equation}
    g^{(2)}_\text{JC} = \frac{\braket{\hat{a}^\dagger \hat{a}^\dagger \hat{a} \hat{a}}_{ss}}{\braket{\hat{a}^\dagger \hat{a}}_{ss}^2},
    \label{eq:g2JC}
\end{equation}
where the steady-state ${\hat{\rho}}^{(\text{JC})}_{ss}$ is obtained from the standard master equation~\cite{breuer2002theory},
\begin{align}
    \nonumber
    \frac{\partial}{\partial t} \hat{\rho}^{(\text{JC})} = &\frac{1}{i \hbar} [\hat{H}_\text{JC}, \hat{\rho}^{(\text{JC})}] + \frac{\Gamma}{2} \mathcal{D}_{\hat{\sigma}^\dagger}[\hat{\rho}^{(\text{JC})}] +\\
    &+ \frac{\gamma}{2} \mathcal{D}_{\hat{\sigma}}[\hat{\rho}^{(\text{JC})}] + \frac{\kappa}{2} \mathcal{D}_{\hat{a}}[\hat{\rho}^{(\text{JC})}],
    \label{eq:MEJC}
\end{align}
as the solution to $\partial_t {\hat{\rho}}^{(\text{JC})}_{ss}=0$.

The JCM summarized in Eqs.~\eqref{eq:HJC} and~\eqref{eq:MEJC} has been extensively used to describe the properties of weakly-coupled CTSs~\cite{carmichael1999statistical}. In this work, it establishes a point of comparison to identify the features that can arise in the QRM.
We use the JCM to calculate the intensity correlations $g^{(2)}_\text{JC}(0)$, shown in dashed orange and blue lines in Fig.~\ref{fig:setup_problem}(b) for the strong and weak incoherent pumping, respectively. In the former case, the JCM correctly reproduces the results of the exact QRM below the USC threshold $\eta\lesssim 0.1$, but fails for larger $\eta$, where the $g^{(2)}(0)$ obtained with the JCM saturates below $g^{(2)}(0)<1$ (see Appendix~\ref{Appendix:JC} for analytical equations), while, in the exact QRM, $g^{(2)}(0)$ keeps increasing with $\eta$. Furthermore, for the weak incoherent pumping (blue lines), we find significant differences between the JCM and QRM even in the WC regime (the JCM and QRM differ for $\eta\gtrsim 5\times 10^{-3}$). This unexpected breakdown of the JCM for small $\eta$ is discussed in detail below.

\section{Origin of the bunching in ultrastrongly coupled systems}
\label{sec:origin.of.bunching}
\begin{figure}[t!]
    \centering
    \includegraphics[width=.9\linewidth]{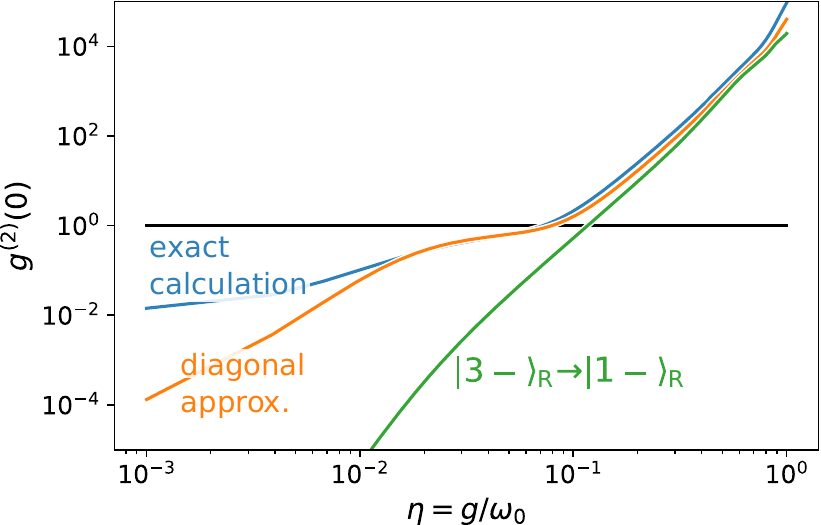}
    \caption{Intensity correlations within several approximations: exact values of $g^{(2)}(0)$ (solid blue line); the diagonal approximation truncated to the states $\{\ket{0}_\text{R}, \ket{1-}_\text{R}, \ket{1+}_\text{R}, \ket{2-}_\text{R}, \ket{2+}_\text{R}, \ket{3-}_\text{R}\}$ in Eq.~\eqref{eq:DiagonalSS} (solid orange line); 
    the diagonal approximation considering only the ${}_\text{R}\!\bra{3-} \hat{x}_a\hat{x}_a \ket{1-}_\text{R}$ term in the numerator of $g^{(2)}(0)$ in Eq.~\eqref{eq:g2_semi} ($\ket{3-}_\text{R} \to \ket{1-}_\text{R}$, solid green line). The calculations shown in this figure are obtained with the QRM for $\Gamma/\gamma = 10^{-3}$.}
    \label{fig:compare_g2s}
\end{figure}

In this Section, we demonstrate that the strong bunching identified in the USC regime in Fig.~\ref{fig:setup_problem}(b) can be traced to characteristics (decay pathways and population) of the single $\ket{3-}_\text{R}$ eigenstate. To justify the focus on that particular eigenstate of the QRM Hamiltonian, in Fig.~\ref{fig:compare_g2s} we plot the exact values of $g^{(2)}(0)$ (solid blue line) obtained for an intermediate pumping $\Gamma/\gamma = 10^{-3}$ together with approximated results. We start by approximating the steady-state density matrix as being diagonal in the basis of the $\ket{\nu}_\text{R}$ eigenstates of the QRM Hamiltonian,
\begin{equation}
\hat{\rho}_{ss} \approx \sum_\nu R_{\nu}~\ket{\nu}\!{}_\text{R}~{}_\text{R}\!\bra{\nu}.
\label{eq:DiagonalSS}
\end{equation}
The approximation is justified by noting that the system is driven with an incoherent pumping mechanism, and -- much like any weakly coupled, thermally populated system -- should exhibit limited coherence. We explore the validity of this approximation in Appendix \ref{Appendix:diagonal.rho}, where we show the values of the steady state density matrix as a function of the coupling strength. We can thus approximate $g^{(2)}(0)$ (Eq.~(\ref{eq:g2QRM})) as (see derivation in Appendix \ref{app:Derive_g2_semi})
\begin{equation}
    g^{(2)}(0) \approx \frac{\sum_{\nu, \mu} R_\nu~ |{}_\text{R}\!\bra{\mu} \hat{x}_a\hat{x}_a \ket{\nu}_\text{R}|^2}{\left(\sum_{\nu, \mu} R_\nu~ |{}_\text{R}\!\bra{\mu} \hat{x}_a \ket{\nu}_\text{R}|^2\right)^2},
    \label{eq:g2_semi}
\end{equation}
where $\ket{\mu}_\text{R}$ and $\ket{\nu}_\text{R}$ are the eigenstates of the QRM Hamiltonian. The intensity correlations calculated by truncating the double sum in the numerator up to $\ket{3-}_\text{R}$ are shown with the solid orange line in Fig.~\ref{fig:compare_g2s} --- this approximation gives a very good agreement with the exact calculation for $\eta \gtrsim 2.5 \times 10^{-2}$; as we have numerically verified, the deviation observed in the WC regime $\eta\lesssim 2.5 \times 10^{-2}$ is not due to the truncation of the basis, but rather due to the effect of the off-diagonal terms of $\hat{\rho}_{ss}$. 

We can further approximate $g^{(2)}(0)$ by by limiting the double sum in Eq.~(\ref{eq:g2_semi}) over $\nu$ and $\mu$ to the $\ket{\nu} = \ket{3-}_\text{R}$ and $\ket{\mu} = \ket{1-}_\text{R}$ term:
\begin{equation}
    g^{(2)}(0) \approx \frac{R_{3-}~ |{}_\text{R}\!\bra{1-} \hat{x}_a\hat{x}_a \ket{3-}_\text{R}|^2}{\left(\sum_{\nu, \mu} R_\nu~ |{}_\text{R}\!\bra{\mu} \hat{x}_a \ket{\nu}_\text{R}|^2\right)^2}.
    \label{eq:g2_semi_last}
\end{equation}
This approximation, denoted in Fig.~\ref{fig:compare_g2s} with the green line, explores the role of the particular, correlated two-photon emission from $\ket{3-}_\text{R}$ to the $\ket{1-}_\text{R}$ state.


\begin{figure}[t!]
    \centering
    \includegraphics[width=\linewidth]{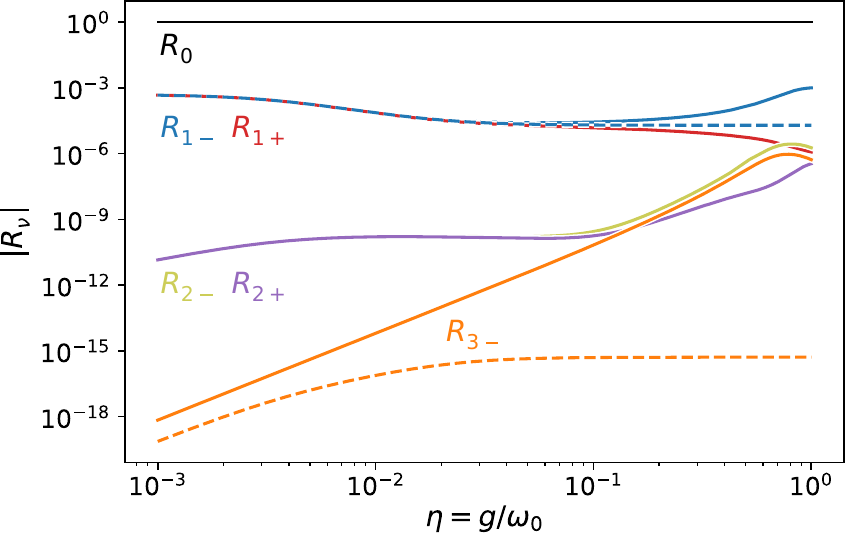}
    \caption{Populations of the polaritonic states $R_{\nu}$ (see labels) calculated with the QRM (solid lines) and JCM (dashed lines), for the intermediate pumping $\Gamma/\gamma = 10^{-3}$.}
    \label{fig:Populations}
\end{figure}

\begin{figure}[t!]
    \centering
    \includegraphics[width=1\linewidth]{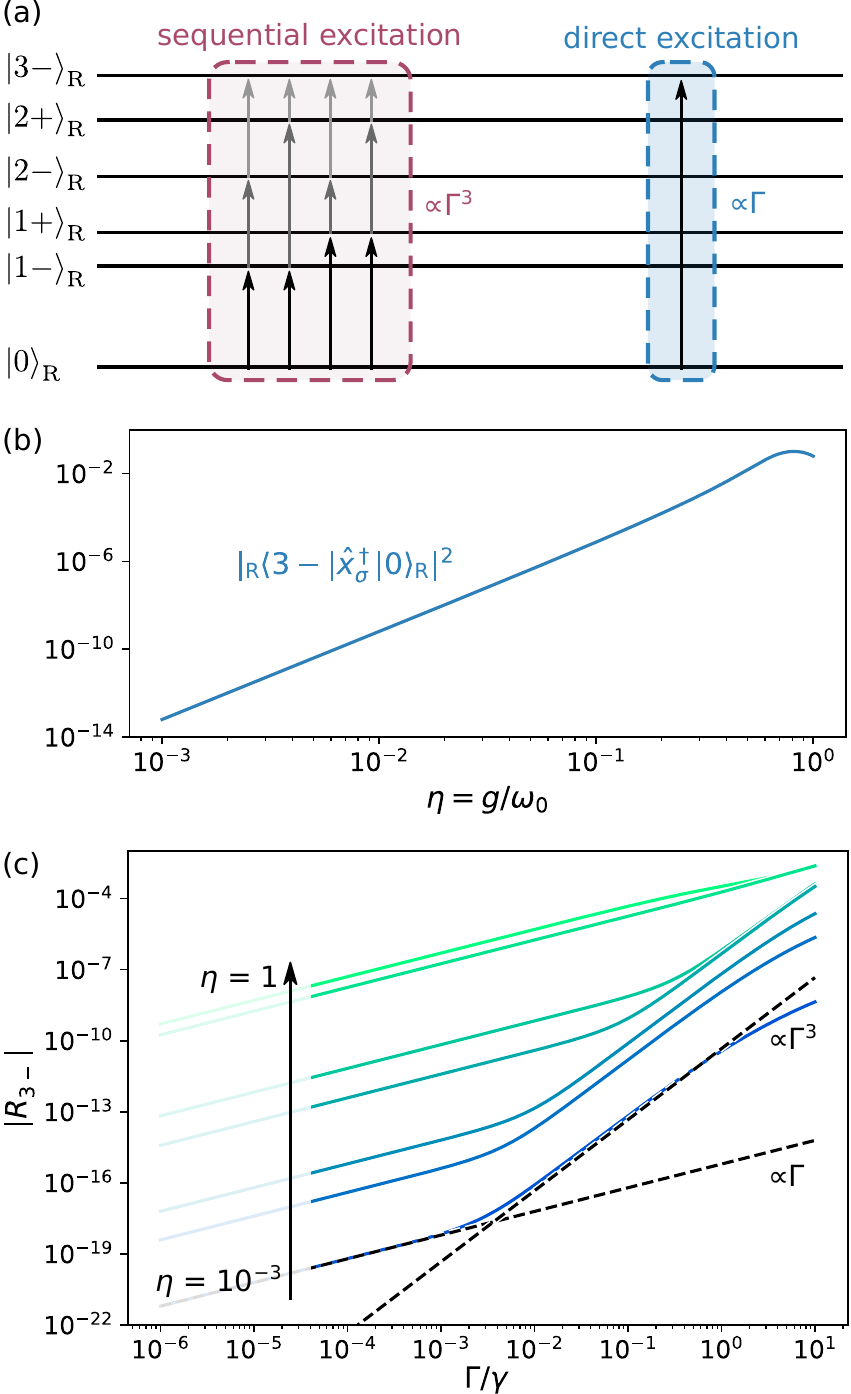}
    \caption{(a) Schematic and (b,c) dependencies of the sequential, and direct excitation pathways of the $\ket{3-}_R$ polariton in the QRM. (a) The two mechanisms are characterised by a cubic and linear dependence of the process on the incoherent driving rate $\Gamma$, respectively. (b) Efficiency of the direct excitation of $\ket{3-}_\text{R}$ state directly from the $\ket{0}_\text{R}$ state, quantified as $|{}_\text{R}\!\bra{3-}\hat{x}_\sigma^\dagger \ket{0}_\text{R}|^2$, growing approximately as $\eta^4$ (until $\eta \approx 0.5$), and matching the dependence of $R_{3-}$ shown in Fig.~\ref{fig:Populations}. (c) Populations of the $\ket{3-}_\text{R}$ eigenstate as a function of incoherent pumping rate $\Gamma$, for a range of coupling parameters $\eta$ (from bottom to top, $\eta = 10^{-3}, 5 \times 10^{-3}, 10^{-2}, 5 \times 10^{-2}, 0.1, 0.5, 1$).}
    \label{fig:Transitions}
\end{figure}

Equation~(\ref{eq:g2_semi_last}) shows that the singular role of the $\ket{3-}_\text{R}$ polariton in the onset of bunching can be ascribed to two effects: the presence of a two-photon emission pathway towards $\ket{1-}_\text{R}$, and a comparatively large, non-thermal population of this state, expressed in our model as $R_{3-}$.

To analyze the latter effect in detail, we plot in Fig.~\ref{fig:Populations} the populations $R_\nu$ of all the relevant eigenstates of the QRM Hamiltonian as a function of the normalized coupling $\eta$ for the intermediate pumping $\Gamma/\gamma = 10^{-3}$. Initially, the eigenstates are populated according to their respective eigenvalues, with the lower-energy states being more populated. While this mechanism holds for $\eta < 0.05$, Fig.~\ref{fig:Populations} shows that for larger coupling the population $R_{3-}$ (orange solid line) rapidly grows at a rate proportional to $\eta^4$, becoming larger than $R_{2+}$ (purple solid line), and eventually approaching $R_{1+}$ (red solid line) for $\eta > 0.1$. This increase of $R_{3-}$ far exceeds that observed within the JCM (see the orange dashed line denoting the population of $\ket{3-}$).

This large population of the $\ket{3-}_\text{R}$ polariton can be attributed to the new, \textit{direct excitation} pathway introduced in the QRM (see Fig.~\ref{fig:Transitions}(a)). In this model, the $\ket{3-}_\text{R}$ polariton can be directly driven from the ground state $\ket{0}_\text{R}$, through the ${\Gamma}/{2} \mathcal{D}_{\hat{x}_\sigma^\dagger}$ incoherent pumping term introduced in the master equation~(\ref{eq:MEQRM}). We illustrate this effect in Fig.~\ref{fig:Transitions}(b), by plotting the $|{}_\text{R}\!\bra{3-}\hat{x}_\sigma^\dagger \ket{0}_\text{R}|^2$ matrix element which quantifies the efficiency of the direct excitation pathway, as a function of $\eta$. We identify a clear scaling with $\eta^4$ until $\eta\approx 0.5$, similarly as in $R_{3-}$.

The direct excitation pathway should be proportional to the incoherent pumping rate $\Gamma$, and compete with the \textit{sequential excitation} pathway $\ket{0,g}_\text{R}\to\ket{1\pm}_\text{R} \to \ket{2\pm}_\text{R} \to \ket{3-}_\text{R}$, with total rate $\propto \Gamma^3$. We can identify this competition in Fig.~\ref{fig:Transitions}(c), by plotting the population $R_{3-}$ as a function of $\Gamma$, for a range of coupling parameters $\eta$. The direct mechanism dominates the pumping for small $\Gamma$, where its linear dependence on the pumping rate makes it more efficient than the sequential pumping mechanism. Only when we increase $\Gamma$, does the latter process become more efficient and we recover $R_{3-} \propto \Gamma^3$ \footnote{We also numerically verify that the populations $R_{2\pm}$ are proportional to $\Gamma^2$ throughout this regime, pointing to a sequential driving mechanism.}. 
The transition from the linear to the cubic dependence on $\Gamma$, or from the direct to sequential excitation mechanisms, shifts towards larger $\Gamma$ as we increase the coupling strength $\eta$. This is because the overall efficiency of the direct excitation, governed by the matrix element $|{}_\text{R}\!\bra{3-}\hat{x}_\sigma^\dagger \ket{0}_\text{R}|^2$, grows rapidly with $\eta$ (see Fig.~\ref{fig:Transitions}(b)).

\section{Probing the breakdown of the RWA}
\label{sec:extent}

\begin{figure}[ht!]
    \centering
    \includegraphics[width=\linewidth]{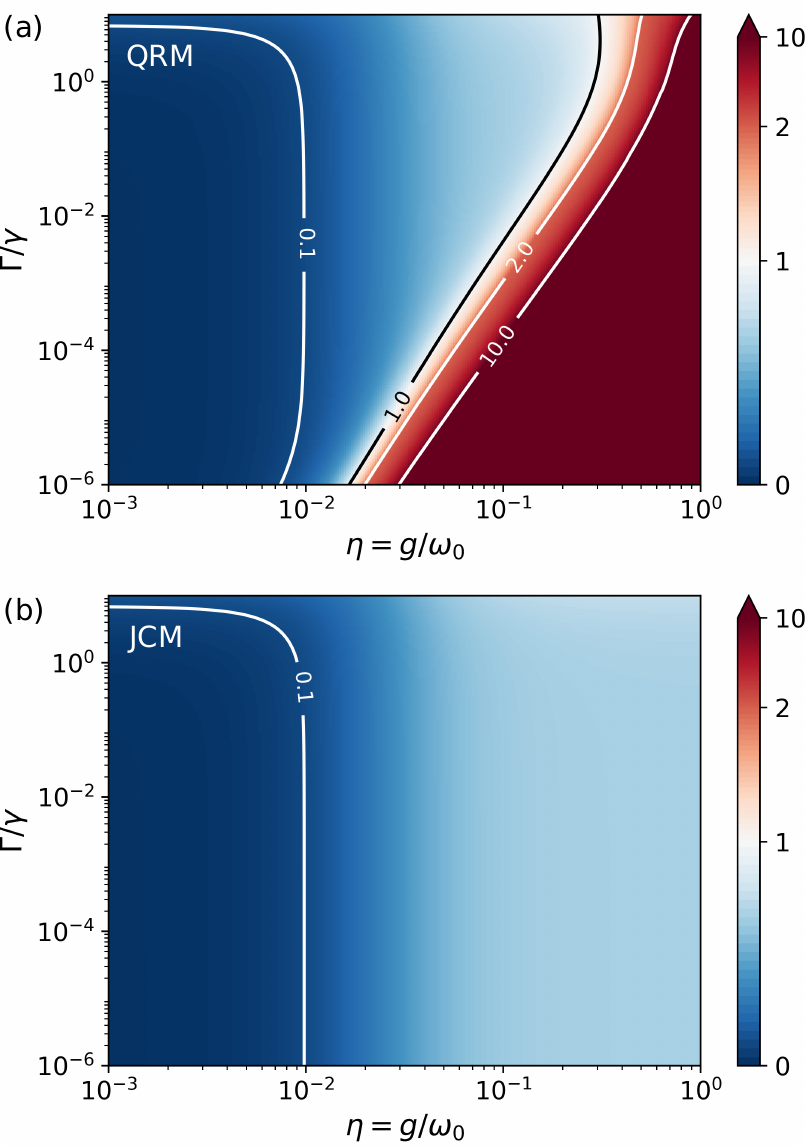}
    \caption{Dependence of $g^{(2)}(0)$ on the pumping rate $\Gamma$ and the coupling strength $\eta$ obtained within the (a) QRM and (b) JCM. The colormap indicates in blue and red the antibunching ($g^{(2)}(0)<1$) and bunching ($g^{(2)}(0)>1$) regions, respectively. The color scale of $g^{(2)}(0)$ is linear from $g^{(2)}(0) = 0$ to $2$ and logarithmic from $g^{(2)}(0) = 2$ to $10$, where it saturates.}
    \label{fig:g2_colormap}
\end{figure}

In Section~\ref{sec:origin.of.bunching} we traced the bunching in the USC regime seen in Fig.~\ref{fig:setup_problem}(b) and Fig.~\ref{fig:compare_g2s} to the direct excitation, and two-photon emission from the $\ket{3-}_\text{R}$ polariton. Here, we explore the extent of this new effect, and identify the threshold for the deviation between the JCM and the QRM. Figures~\ref{fig:g2_colormap}(a) and (b) compare the values of $g^{(2)}(0)$ calculated with the QRM and the JCM, respectively, for a range of the normalized coupling strengths $\eta$ and pumping rates $\Gamma$, and demonstrate that the deviation between JCM and QRM also depends strongly on the latter parameter ($\Gamma$), in a manner that can be explained using the formulation laid out in Section~\ref{sec:origin.of.bunching}.

\begin{figure}[ht!]
    \centering
    \includegraphics[width=\linewidth]{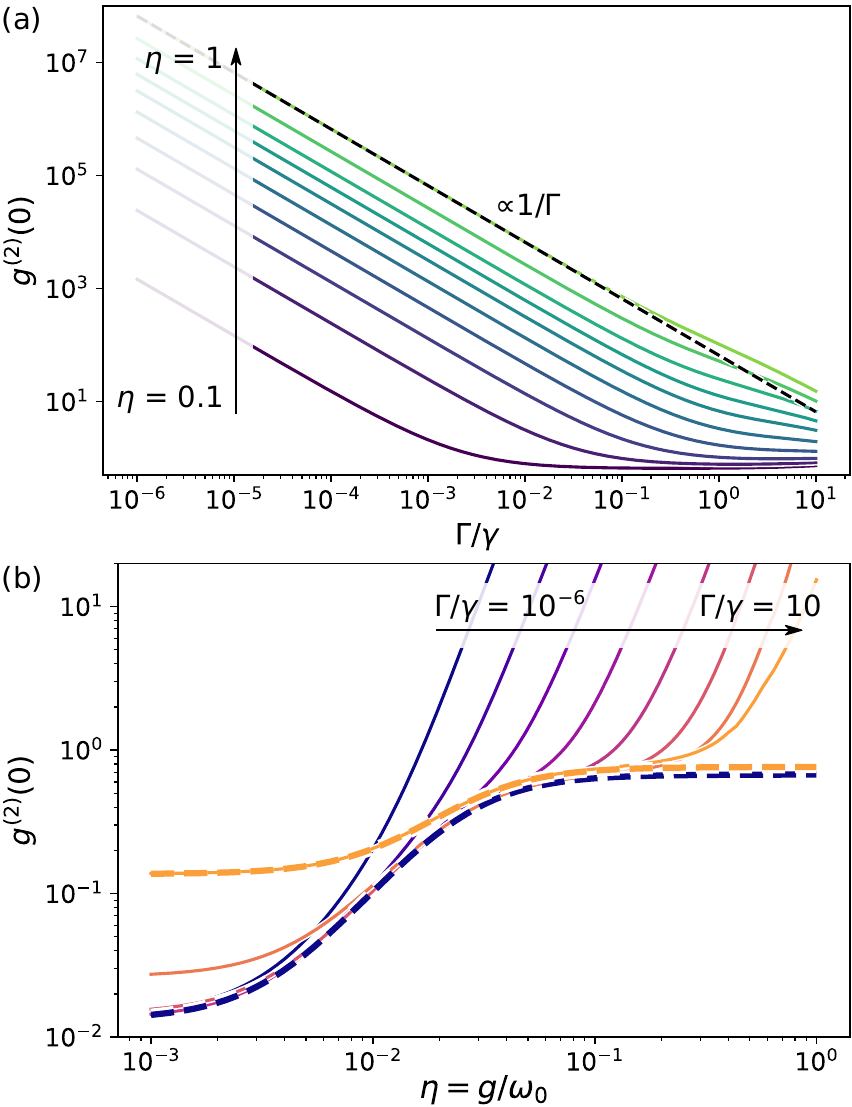}
    \caption{Landscape of $g^{(2)}(0)$ as a function of (a) normalized incoherent pumping rate $\Gamma/\gamma$, and (b) coupling parameter $\eta=g/\omega_0$. In (a) the collection of lines represents results for the coupling parameter varied linearly in the range $\eta \in [0.1, 1]$. In (b) we exponentially increase the pumping rates from $\Gamma/\gamma=10^{-6}$ to $10$. The solid lines are obtained with the QRM, and the dashed orange and blue lines in (b) show the results obtained within the JCM for $\Gamma/\gamma=10^{-6}$ and $10$, respectively.}
    \label{fig:g2_P}
\end{figure}

\subsection{Qualitative dependence of the bunching on the pumping}
\label{subsec:population_dependence}

The map of intensity correlation shown in Fig.~\ref{fig:g2_colormap}(a) indicates that the bunching observed in the QRM depends on the rate of  incoherent pumping $\Gamma$ for a wide range of coupling parameters $\eta$. We explore this effect in more detail in Fig.~\ref{fig:g2_P}(a), where we plot vertical cross sections of Fig.~\ref{fig:g2_colormap}(a) --- the dependence of $g^{(2)}(0)$ on $\Gamma$ ---, for $\eta=0.1$ to $1$, and find that the intensity correlations follow
\begin{equation}\label{eq:g2.Gamma}
    g^{(2)}(0) \propto \frac{1}{\Gamma}.
\end{equation}
This dependence results from the direct excitation mechanism of the polariton $\ket{3-}_\text{R}$: In the USC regime, the contribution from that polariton dominates the numerator of $g^{(2)}(0)$ approximated as in Eq.~\eqref{eq:g2_semi_last}, with the numerator proportional to $R_{3-} \propto \Gamma$ (as discussed in section~\ref{sec:origin.of.bunching}). Conversely, the denominator of Eq.~\eqref{eq:g2_semi_last} is dominated by the contribution from the $\ket{1\pm}_\text{R}$ eigenstates $\propto R_{1\pm}$. Since both $\ket{1-}_\text{R}$ and $\ket{1+}_\text{R}$ are populated directly from the ground state, we find $R_{1\pm} \propto \Gamma$. Thus we recover the $g^{(2)}(0) \propto R_{3-}/(R_{1\pm})^2 \propto 1/\Gamma$ dependence, and $g^{(2)}(0)$ is therefore unbounded.

Equation~(\ref{eq:g2.Gamma}) can be used as a marker for the breakdown of the JCM, since in that model, the direct pumping mechanism of high-order polaritons is absent, and consequently the intensity correlations are largely constant for small $\Gamma$ (see Fig.~\ref{fig:g2_colormap}(b)). This is further explored in the following subsection. 

\subsection{Breakdown of the JCM Hamiltonian in the WC regime}
\label{subsubsec:breakdown}

Figure~\ref{fig:g2_colormap}(a) shows that in the QRM, the strong bunching region can also be identified for $\eta$ in the WC and SC regimes. In particular, for very weak incoherent pumping ($\Gamma / \gamma \lesssim 10^{-5}$), the strong bunching appears for  couplings as small as $\eta \approx 2.5 \times 10^{-2}$, highlighting a significant deviation between the predictions of the JCM (where $g^{(2)}<1$ for all $\eta$) and of the QRM. To demonstrate this effect more clearly, in Fig.~\ref{fig:g2_P}(b) we plot horizontal cross sections of Fig.~\ref{fig:g2_colormap}  --- the dependence of $g^{(2)}(0)$ on $\eta$ calculated for different pumping rates $\Gamma$ using the QRM (solid lines) and in the JCM (dashed lines) models.

The difference between the QRM and JCM in the WC and SC regimes has the same origin as in the USC regime: the direct excitation of the $\ket{3-}_\text{R}$ eigenstate in the QRM. As we discuss in Section~\ref{sec:origin.of.bunching}, this direct mechanism can be dominant for any $\eta$, providing that the pumping is sufficiently small. Conversely, in the JCM, for small $\Gamma$ ($\Gamma \ll \gamma$), $g^{(2)}(0)$ is determined by the population of the polariton with two-photon terms $\ket{2\pm}$ (we neglect the far smaller contribution from $\ket{3\pm}$). Since this polariton is excited through a sequential process, its population is proportional to $\Gamma^2$. Normalized by the square of the population of the $\ket{1\pm}$ states ($\propto \Gamma^2$), the intensity correlation in the weak pumping and coupling limit of the JCM is independent of $\Gamma$, and does not support bunched emission pathways (we derive this result explicitly in Appendix~\ref{Appendix:JC}).


It is worthwhile to consider how these striking differences can emerge in the WC limit, and far below the USC threshold $g \ll 0.1 \omega_0$, where the QRM and JCM would be conventionally expected to match. To derive $\hat{H}_\text{JC}$, in Eq.~\eqref{eq:HJC} we (i) performed the expansion of the QRM Hamiltonian $\hat{H}$ in a power series of $\eta$, and dropped terms scaling with higher powers of $\eta$, and (ii) applied the RWA to remove the non-number-conserving terms. Figure~\ref{fig:g2_colormap} shows that QRM and JCM can diverge for arbitrarily small $\eta$, suggesting that the error in $g^{(2)}(0)$ introduced by the series truncation in (i) can be made arbitrarily small. Thus, we can trace the observed differences to the application of the RWA. 



This is, to our knowledge, the first proposal for observing the breakdown of the RWA in the WC limit. An experimental assessment of this effect would constitute a challenge, mostly due to the very short decoherence time of the molecules, further reduced by the high-cooperativity coupling to the cavities, resulting in sub-ps timescales for the detection of bunching (see discussion in Appendix~\ref{app:experimental}). These requirements can be partially relaxed by considering atoms with smaller $\gamma$, and reduced cooperativity. Alternatively, a weaker pumping - while reducing the emission and coincidence rate - should boost the bunching effect (see Fig.~\ref{fig:g2_colormap}), making its observation possible even after averaging over the longer detection window.


\subsection{Emission spectra}
\label{subsec:emission_spectra}


To conclude our characterization of the QRM, we now study the dependence of the one-photon emission spectrum $S^{(1)}(\omega)$ on the coupling strength. We show that the differences below the USC between the QRM and the JCM are much smaller for $S^{(1)}(\omega)$ than for the intensity correlations $g^{(2)}(0)$, emphasizing the interest of using the intensity correlation to identify the breakdown of the JCM.



\begin{figure}[t!]
    \centering
    \includegraphics[width=\linewidth]{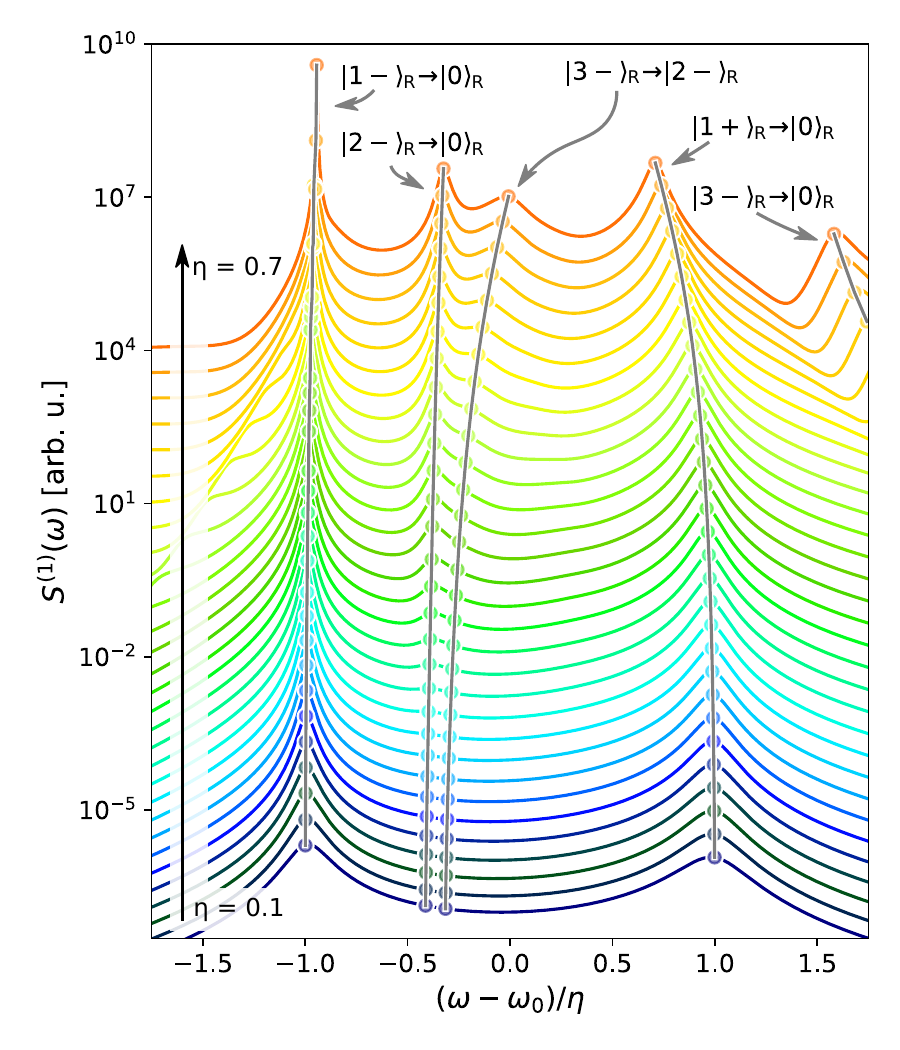}
    \caption{Emission spectra from the USC system for the parameter $\eta$ changing linearly from 0.1 to 0.7, in $\Delta \eta = 0.02$ steps. The spectra are shifted vertically for clarity. Features describing transitions between eigenstates of the QRM Hamiltonian are traced and marked. The spectra are calculated within the QRM for an intermediate pumping rate $\Gamma / \gamma = 10^{-3}$.}
    \label{fig:spectrum}
\end{figure}

In Fig.~\ref{fig:spectrum}, we plot the emission spectra $S^{(1)}(\omega)$ for a range of coupling strengths $\eta$. We calculate $S^{(1)}(\omega)$ within the QRM as~\cite{Hughes2022gauge, breuer2002theory, PhysRevResearch.3.023079}
\begin{equation}
    S^{(1)}(\omega) \propto \int_{-\infty}^{\infty} \braket{\hat{x}_a^\dagger (\tau) \hat{x}_a(0)}_{ss}e^{-i\omega \tau} d\tau,
    \label{eq:EmissionSpectrum}
\end{equation}
where $\hat{x}_a$ and $\hat{x}_a^\dagger$ are the dressed annihilation and creation operators of the cavity, respectively (see Eq.~\eqref{eq:dressed_a}). {For a detailed study of the emission from the TLS ($S^{(1)}_\sigma$ calculated from the two-time correlators between $\hat{x}_\sigma$ operators) in the USC regime, we direct the reader to Ref.~\cite{PhysRevResearch.4.023048}.} 
Each spectral feature in the figure corresponds to a transition between the eigenstates of the system, which we identify by matching the frequency of the peak with the difference between the eigenfrequencies of the system (see Fig.~\ref{fig:Eigenvalues}(b)). We plot these spectra as functions of the rescaled frequency $(\omega-\omega_0)/\eta$ to compensate for the $\eta$-dependence of the Rabi splitting between the polaritons.

The smallest coupling strength considered in Fig.~\ref{fig:spectrum} corresponds to the low limit of the USC, $\eta = 0.1$. This spectrum displays two peaks of similar intensity which, as in the JCM, correspond to the emission of a single photon via the $\ket{1-}_\text{R}\to \ket{0}_\text{R}$ (lower frequency feature) and $\ket{1+}_\text{R}\to \ket{0}_\text{R}$ (higher frequency) transitions. Only when we increase the coupling strength to about $\eta\approx 0.3$, do the spectra develop additional features: the strength of the spectral peak corresponding to the $\ket{1-}_\text{R} \to \ket{0}_\text{R}$ transition increases, and two new peaks, corresponding to the $\ket{3-}_\text{R} \to \ket{2-}_\text{R}$ and $\ket{2-}_\text{R} \to \ket{0}_\text{R}$ transitions (as labelled in the figure), emerge. The visibility of each peak can be compared to the populations of the initial states participating in the emission processes. For instance, the two new peaks follow the same dependence on $\eta$ as the populations $R_{3-}$ and $R_{2-}$, respectively (Fig.~\ref{fig:Populations}), so that, for $\eta > 0.05$, the two populations begin to grow very rapidly with increasing $\eta$. Simultaneously, as we increase the coupling $\eta$, spectral features continuously shift, reflecting the changes to the spectrum of the QRM Hamiltonian (Fig.~\ref{fig:Eigenvalues}(b)).

\begin{figure}[t!]
    \centering
    \includegraphics[width=\linewidth]{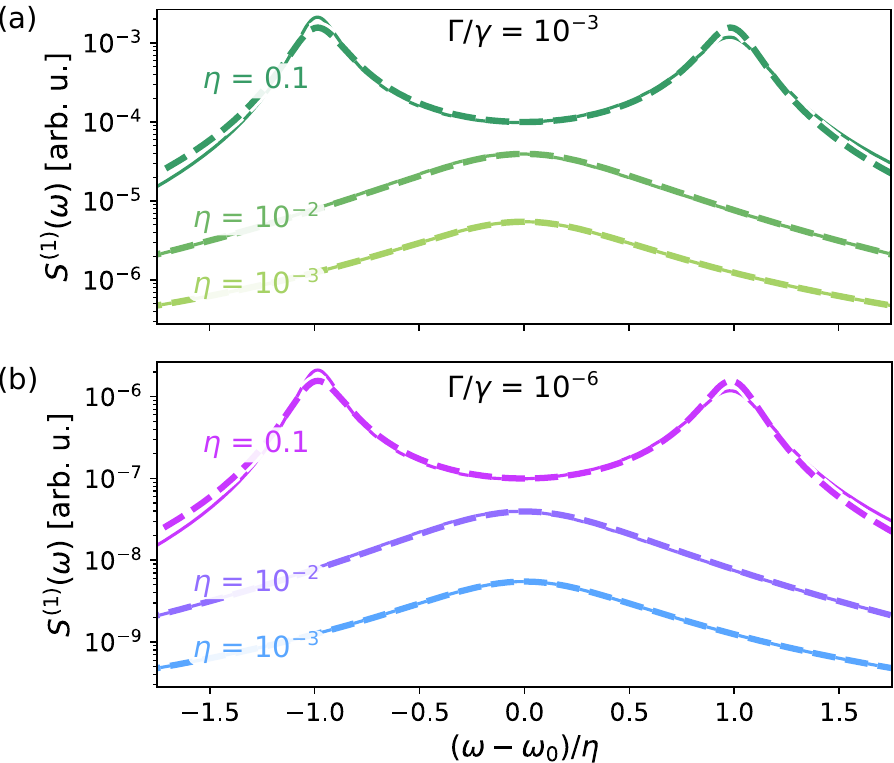}
    \caption{Comparison between the emission spectra obtained with the QRM (solid lines) and JCM (dashed lines) for different couplings strengths, as indicated in the figure. Panels (a) and (b) correspond to different pumping rate, $\Gamma / \gamma = 10^{-3}$ and $\Gamma / \gamma = 10^{-6}$, respectively. }
    \label{fig:spectrum_QRM_vs_JCM}
\end{figure}

These additional peaks emerge only in the USC regime, for $\eta \gtrsim 0.3$. To check whether this threshold can be significantly modified, as we observed for the intensity correlations, in Fig.~\ref{fig:spectrum_QRM_vs_JCM} we compare the spectra obtained from the JCM (dashed lines) and the QRM (solid lines) using weaker incoherent pumping $\Gamma$, and various couplings $\eta = 10^{-3}$, $10^{-2}$, and $0.1$ (note that $\eta = 10^{-3}$ and $10^{-2}$ are below the minimum value considered in Fig.~\ref{fig:spectrum}, and give no peak-splitting). The spectra point to small differences between the two models for the larger $\eta$ for both considered pumping rates, but these differences would be likely difficult to identify in  experimental settings. On the other hand, as shown in Figs. \ref{fig:g2_colormap} and \ref{fig:g2_P}(b), the $g^{(2)}(0)$ obtained for $\Gamma / \gamma = 10^{-3}$ and $\Gamma / \gamma = 10^{-6}$ show qualitative differences for $\eta \gtrsim 2.5 \times 10^{-2}$ due to the direct excitation pathway introduced in Section \ref{sec:origin.of.bunching}. The lack of sensitivity of $S^{(1)}(\omega)$ to the direct excitation pathway is due to the fact that the emission from the lower $\ket{1-}_\text{R}$ and $\ket{1+}_\text{R}$ states predominantly govern the emission spectra for $\eta\lesssim0.3$. The excitation and emission from these eigenstates are not impacted by the direct excitation mechanism discussed above, and do not flag the breakdown of the RWA. We thus conclude that the characterization of the correlations is a more powerful tool than measuring the one-photon emission spectra for the identification of phenomena caused by the non-number-conserving terms of the QRM Hamiltonian for coupling below the traditional USC threshold $\eta>0.1$.

\section{Conclusions}

In this work, we analyze the statistics of the emission from a generic quantum system comprising an incoherently driven two-level emitter interacting with a cavity. We identify an emergence of unbounded bunching as the system approaches the USC regime. By expressing the dynamics of the system in the basis of the polaritonic eigenstates of the quantum Rabi Hamiltonian, we can attribute the bunching to the singular behavior of the individual eigenstate $\ket{3-}_\text{R}$, which (i) decays through a correlated two-photon emission, and (ii) is very strongly populated by a new, direct excitation mechanism from the ground state. 

Our work shows that intensity correlations $g^{(2)}(0)$ are a much more sensitive tool for observing the phenomena induced by the non-number-conserving terms in the QRM, than the one-photon emission spectra. Indeed, we find that the intensity correlations can identify a breakdown of the rotating wave approximation far below the conventional limit of the USC, with the exact limit determined by the rate of incoherent pumping. 

These findings calls for an experimental verification, and further theoretical studies, to verify the robustness of the identified excitation and emission mechanisms. Our model can be extended to account for the more complex decay dynamics and energy structure of the quantum emitter, involving dark excitonic states, or pure dephasing, as well as the interaction with a structured reservoir.

\section*{Acknowledgements}
{Á. N. and M.~K.~S. thank A. Gonz\'alez-Tudela and D. Burgarth for their useful inputs and valuable discussions.} Á. N., R. E., U. M., and J. A. acknowledges the financial support from the Spanish Ministry of Science and Innovation and the Spanish government agency of research MCIN/AEI/10.13039/501100011033 through Project Ref. No. PID2019-107432GB-100, and from the Department of Education, Research and Universities of the Basque Government through Project Ref. No. IT 1526-22. M.K.S. acknowledges funding from the Macquarie University Research Fellowship Scheme (MQRF0001036), and the Australian Research Council Discovery Early Career Researcher Award DE220101272.

\bibliography{references}


\appendix

\section{Correlations in the Jaynes-Cummings Hamiltonian}
\label{Appendix:JC}

In this Appendix, we derive an analytical expression for $g^{(2)}(0)$ as obtained with the approximated JCM Hamiltonian. We focus here on the same CTS system introduced in the main text, but we consider only the weak-illumination case, where the incoherent pumping of the TLS has a rate much smaller than the TLS losses, $\Gamma \ll \gamma$.
We identify the minimum set of operators for which the master equations form an almost closed system (see discussion below): $\mathbf{v} = 
\left(\hat{a}^\dagger\hat{a}, 
\hat{\sigma}^\dagger \hat{\sigma}, 
\hat{a}^\dagger\hat{\sigma}, 
\hat{a}\hat{\sigma}^\dagger, \hat{a}^\dagger\hat{a}\hat{\sigma}^\dagger\hat{\sigma},
\hat{a}^\dagger\hat{a}\hat{a}\hat{\sigma}^\dagger,
\hat{a}^\dagger\hat{a}^\dagger\hat{a}\hat{\sigma},
\hat{a}^\dagger\hat{a}^\dagger\hat{a}\hat{a}\right)^T$.
The equations of motion for the expectation values of these operators can be approximately expressed as
\begin{equation}
    \frac{d}{dt} \braket{\mathbf{v}} = M\braket{\mathbf{v}} + \mathbf{b},
\end{equation}
with 
\begin{widetext}
\begin{equation}
    M = \begin{pmatrix}
    - \kappa & 0 & -g & -g & 0 & 0 & 0 & 0\\
    0 & - \gamma & g & g & 0 & 0 & 0 & 0\\
    g & -g & -\frac{1}{2}(\Gamma + \kappa + \gamma) & 0 & -g & 0 & 0 & 0\\
    g & -g & 0 & -\frac{1}{2}(\Gamma + \kappa + \gamma) & -g & 0 & 0 & 0\\
    \Gamma & 0 &0 & 0 & -(\gamma + \kappa) & g & g & 0 \\
    0 & 0 & 0 & 0 & -2g & -\frac{1}{2}(\Gamma+\gamma+3 \kappa) & 0 & g\\
    0 & 0 & 0 & 0 & -2g & 0 & -\frac{1}{2}(\Gamma+\gamma+3 \kappa) & g\\
    0 & 0 & 0 & 0 & 0 & -2g & -2g & -2 \kappa
    \end{pmatrix},
\end{equation}
\end{widetext}
and vector $\mathbf{b} = (0,\Gamma,0,0,0,0,0,0)^T$. Here we truncated the set of equations by considering that the pumping rate of the system, $\Gamma$, is very small. Thus, in the steady-state, the TLS is mostly in the ground state. Hence, we approximate~\cite{del2009luminescence}: (i) $\braket{\hat{\sigma}\hat{\sigma}^\dagger} \approx 1$, (ii) $\braket{\hat{a}^\dagger\hat{a}^\dagger \hat{a} \hat{a} \hat{\sigma} \hat{\sigma}^\dagger} \approx \braket{\hat{a}^\dagger\hat{a}^\dagger \hat{a} \hat{a}}$, and (iii) $\braket{\hat{a}^\dagger\hat{a}^\dagger \hat{a} \hat{a} \hat{\sigma}^\dagger \hat{\sigma}} \approx 0$. In the steady state $\frac{d}{dt}\braket{\mathbf{v}}_{ss}=0$, and we can derive closed expressions for $\braket{\hat{a}^\dagger \hat{a}}_{ss}$ and $\braket{\hat{a}^\dagger\hat{a}^\dagger\hat{a} \hat{a}}_{ss}$. Instead of writing their full forms, we 
are interested only in the values of $\braket{\hat{a}^\dagger \hat{a}}_{ss}$ and $\braket{\hat{a}^\dagger\hat{a}^\dagger\hat{a} \hat{a}}_{ss}$ in the two opposite limits of $g\ll \Gamma$ (vanishing coupling), and $g\gg \kappa$ (beyond the SC regime). We obtain an expression that is valid in both limits by retaining the terms dependent on the leading powers of the two free parameters $\Gamma$ and $g$: 
\begin{equation}
    \braket{\hat{a}^\dagger \hat{a}}_{ss} \approx \frac{4 g^2 (4 g^2  + \kappa^2) \Gamma}{\kappa^3\left(4 g^2+\gamma \kappa\right)}
    =\Gamma \frac{C}{C+1}\frac{4 g^2  + \kappa^2}{\kappa^3},
    \label{eqS1A}
\end{equation}
where $C=4g^2/(\kappa\gamma)$ is the cooperativity, and
\begin{align}
    \braket{\hat{a}^\dagger\hat{a}^\dagger\hat{a} \hat{a}}_{ss} \approx \frac{32 g^4 \Gamma^2}{3\kappa^2(16 g^4 + \gamma \kappa^3)}.
    \label{eq:S2A}
\end{align}
The $\Gamma^2$ dependence of $\braket{\hat{a}^\dagger\hat{a}^\dagger\hat{a} \hat{a}}_{ss}$ reflects our intuition that the two-photon states in the JCM should be excited via a sequential excitation from the ground state. This intuition fails for the QRM Hamiltonian, as discussed in Section~\ref{sec:extent}. 

The intensity correlations are found, in the two limits of interest, as 
\begin{equation}
    g^{(2)}(0) \xrightarrow{\kappa \gg \gamma \gg \Gamma \gg g} \frac{2}{3}\frac{\gamma}{\kappa}, 
    \label{eq:g2VanishCoup}
\end{equation}
\begin{equation}
    g^{(2)}(0) \xrightarrow{g\gg \kappa \gg \gamma \gg \Gamma} \frac{2}{3}.
\end{equation}
Note that $g^{(2)}$ here is derived by neglecting the direct emission from the TLS, which is an invalid approximation for very small $g$ in equation \eqref{eq:g2VanishCoup} (unless the emission of the TLS is filtered-out). If direct emission of the TLS is included \eqref{eq:g2VanishCoup} should be modified. The more clear situation is in the limit of $g=0$. In this case, the emission of the system is only given by the TLS, which emits one photon at a time resulting in $g^{(2)}=0$.

\section{Approximating the steady-state density matrix}
\label{Appendix:diagonal.rho}

\begin{figure*}[ht!]
    \centering
    \includegraphics[width=\textwidth]{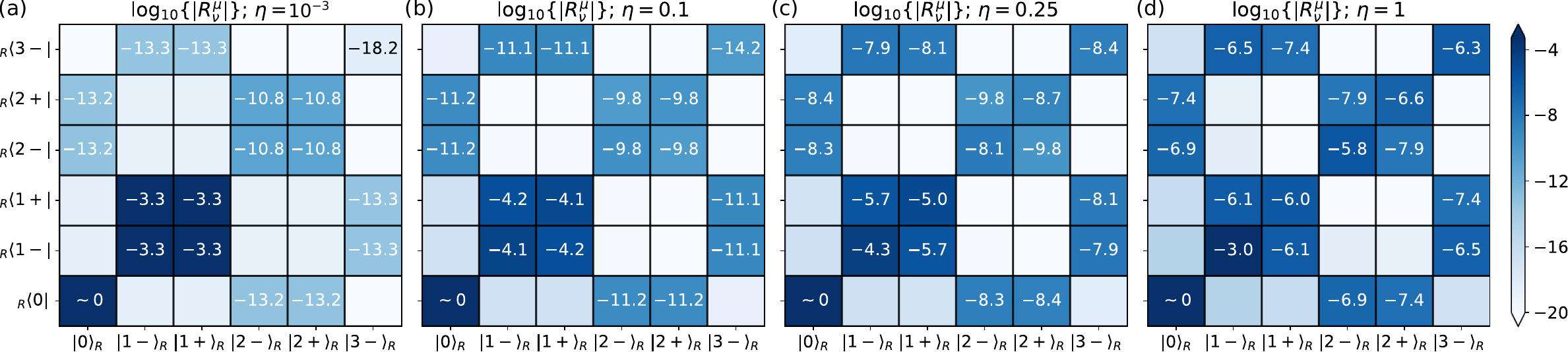}
    \caption{(a)-(d) Elements of the steady-state density matrices calculated for increasing values of the coupling parameter $\eta=g/\omega_0$ denoted at the top. The figure shows the logarithm of the absolute value of the first $6\times 6$ elements of the density matrices (see Eq.~\eqref{eq:Rmn}, where the columns and rows in the figure indicate the different combinations of $\ket{\mu}\!{}_\text{R}$ and ${}_\text{R}\!\bra{\nu}$ in Eq.~\eqref{eq:Rmn}). The results are color-coded, and the approximated numerical values for the largest terms are explicitly listed in the panels. The density matrices are obtained within the QRM for an intermediate pumping rate, $\Gamma/\gamma = 10^{-3}$.}
    \label{fig:Rij}
\end{figure*}

In Section~\ref{sec:origin.of.bunching}, we calculate $g^{(2)}(0)$ considering several approximations of the density matrix of the steady state, such as truncating the Hilbert space to the $\ket{3-}_\text{R}$ state, and neglecting the off-diagonal elements. We found that these approximations are convenient for analyzing $g^{(2)}(0)$ beyond the WC regime (i.e., for $\eta>2.5 \times 10^{-2}$, see discussion of Fig.~\ref{fig:compare_g2s}). However, the steady-state off-diagonal elements can be crucial in some cases, like evaluating the QRM in the WC regime (see Fig.~\ref{fig:compare_g2s}).

In Fig.~\ref{fig:Rij} we show the diagonal and off-diagonal $R^{\mu}_{\nu}$ elements of the steady-state density matrix,
\begin{equation}
    \hat{\rho}_{ss} = \sum_{\mu, \nu} R^{\mu}_\nu \ket{\mu}\!{}_\text{R}~{}_\text{R}\!\bra{\nu},
    \label{eq:Rmn}
\end{equation}
where $\ket{\mu}_\text{R}$ and $\ket{\nu}_\text{R}$ are the eigenstates of the QRM Hamiltonian. In the figure we show $R^{\mu}_{\nu}$ for the first six eigenstates of the system, $\ket{\mu}_\text{R}, \ket{\nu}_\text{R} = \{\ket{0}_\text{R}, \ket{1-}_\text{R}, \ket{1+}_\text{R}, \ket{2-}_\text{R}, \ket{2+}_\text{R}, \ket{3-}_\text{R}\}$, and four coupling parameters $\eta = 0.001$, $0.1$, $0.25$, and $1$ (as marked in the figure).

All the density matrices show an approximate block-diagonal form, with dominant terms $R^{n\pm}_{n\pm}$ and $R^{n\pm}_{n\mp}$ terms. For the larger couplings (Fig.~\ref{fig:Rij}(b)-(d)), $\hat{\rho}_{ss}$ shows some additional coherence terms between eigenstates with the same parity of excitations, \textit{i.e.}, $R^{m\pm}_{n\pm}$ and $R^{m\pm}_{n\mp}$ terms with $m$ and $n$ ($m\neq n$) being both even or odd numbers (\textit{e.g.}, the $R^{3-}_{1-}$ and $R^{1+}_{3-}$ terms). As we increase $\eta$, the density-matrices experience three main changes: (i) the ratio between the diagonal $R_{n\pm}^{n\pm}$ and off-diagonal $R_{n\mp}^{n\pm}$ terms ($R_{n\pm}^{n\pm} / R_{n\mp}^{n\pm}$) decrease, (ii) the $R_{n-}^{n-}$ population of the lower $\ket{n-}_\text{R}$ eigenstates becomes higher than the $R_{n+}^{n+}$ population of the upper $\ket{n+}_\text{R}$ eigenstates, (iii) the population of the higher order eigenstates increases significantly. In particular, the population of $\ket{3-}_\text{R}$ increases from $R_{3-}^{3-}\approx 10^{-17}$ at $\eta = 0.001$ to $R_{3-}^{3-}\approx 10^{-6}$ at $\eta=1$, and becomes comparable to that of the lowest $n= 1$ eigenstates.


\section{Effective thermal pumping}
\label{Appendix:ThermalPumping}


While our work focuses on the experimentally-viable mechanism of incoherent driving of the TLS, it is worth comparing this framework to other driving mechanisms. In particular, we can consider the pumping of both the TLS and the cavity due to the coupling with two thermal baths, both at the same temperature, as discussed in Refs.~\citenum{ridolfo2013nonclassical} and \citenum{Chen_2022}. In this case, the steady state of the system is given by the statistical mixture of the eigenstates of the QRM Hamiltonian, with parameters
\begin{equation}
    R_\nu{}^\text{(Th.)} \propto \left[\exp\left(\frac{E_\nu}{k_B T}\right)-1\right]^{-1},
    \label{eq:ThermalPop}
\end{equation}
where $E_\nu$ is the energy of the $\nu$th eigenstate (calculated with respect to the energy of the ground state), $k_B$ is the Boltzmann constant, and $T$ is the temperature of the thermal bath. 

\begin{figure}[ht!]
    \centering
    \includegraphics[width=\linewidth]{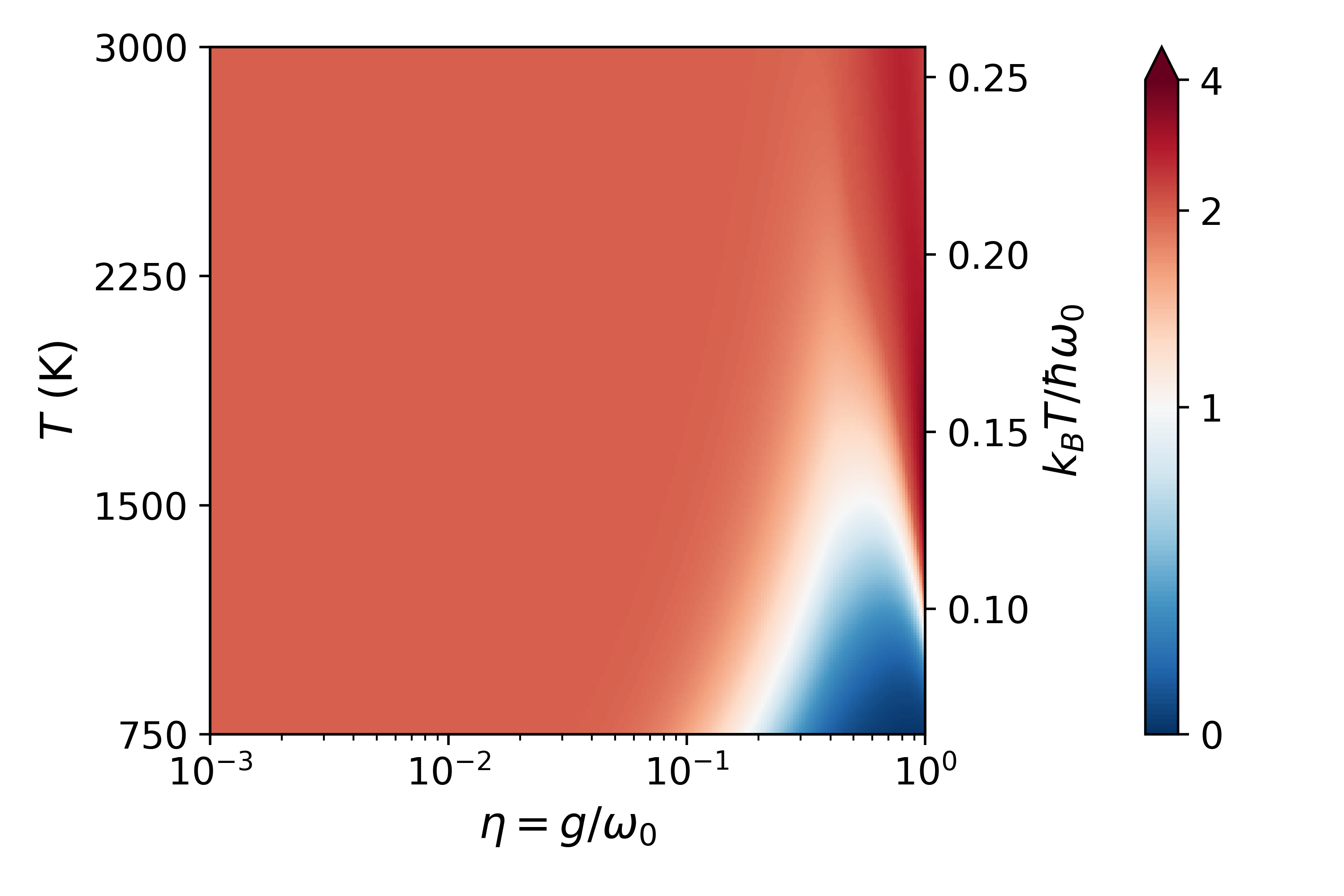}
    \caption{$g^{(2)}(0)$ of a thermally pumped CTS as a function of the coupling strength $\eta$, and of the temperature $T$ of the thermal bath (for reference we give a secondary vertical axis with the values of $k_B T/(\hbar \omega_0)$). The results of the calculation are obtained within the QRM. The colormap indicates in blue and red the antibunching ($g^{(2)}(0)<1$) and bunching ($g^{(2)}(0)>1$) regions, respectively. The colorscale of $g^{(2)}(0)$ is linear from $g^{(2)}(0) = 0$ to $2$ and logarithmic from $g^{(2)}(0) = 2$ to $4$, where it saturates.}
    \label{fig:g2_thermal}
\end{figure}

Figure~\ref{fig:g2_thermal} shows the correlations obtained with the QRM considering the pumping by a thermal bath at different temperatures (from $T = 750$ K to $T = 3000$ K, for reference we also include a secondary vertical axis in the figure with $k_B T / (\hbar \omega_0)$ values). For each temperature, we show the dependence of $g^{(2)}(0)$ on the coupling strength $\eta$. For $\eta \lesssim 0.1$, Fig.~\ref{fig:g2_thermal} shows that regardless of the coupling strength between the cavity and the TLS, the emission of the system follows a thermal statistic, $g^{(2)}(0) = 2$ \cite{ridolfo2013nonclassical}. On the other hand, for $\eta \gtrsim 0.1$, the system emission becomes non-classical: For low temperatures $T\lesssim 1750$ K and large couplings $\eta \gtrsim 0.1$, the system emission results in antibunching with $g^{(2)}(0)<1$. For high temperatures, $T\gtrsim 1750$ K and large couplings $\eta \gtrsim 0.1$, the emission of the system results in a strong bunching with $g^{(2)}(0)>2$. This behavior is very different to the one studied in the main text for incoherent illumination of the TLS. The intensity correlation under thermal pumping in Fig. \ref{fig:g2_thermal} attains a maximum value of 4 (compared to $\approx 10^7$ in Fig. \ref{fig:g2_P}) and it does not show any signature of the unbound increase for weaker pumping. 

The general behavior we show in Fig.~\ref{fig:g2_thermal} is in qualitative agreement with the results reported in Ref.~\cite{ridolfo2013nonclassical}, where the authors also analyze the emission statistics of a thermally pumped CTS as a function of temperature and coupling strength. There are, however, some qualitative differences due to the description of the Hamiltonian and emission operators in the QRM~\cite{Hughes2022gauge, settineri2018dissipation, di_stefano2019resolution} (Eqs. \eqref{eq:g2QRM}-\eqref{eq:dressed_a}). For example, neglecting the Gauge correction of the QRM Hamiltonian result in an antibunched emission of the system for higher temperatures than in Fig.~\ref{fig:g2_thermal} (results not shown here).

\section{Other detection schemes}
\label{app:additional}

\begin{figure}[t]
    \centering
    \includegraphics[width=\linewidth]{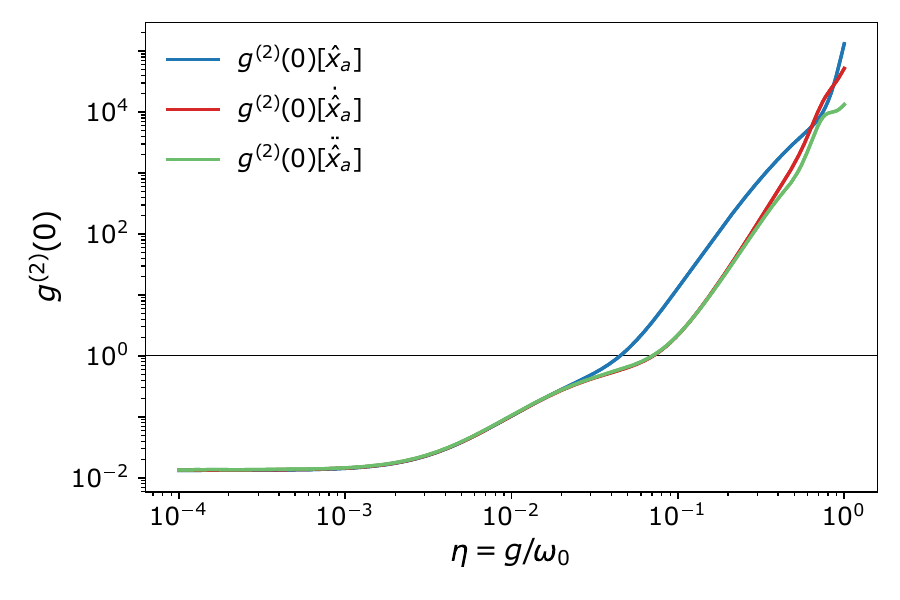}
    \caption{Intensity correlations calculated with different definitions of the intensity correlations. The blue line corresponds to the scenario where the detectors couple to the dressed operators of the photonic excitations in the cavity, $\hat{x}_a$ (see Eq.~\eqref{eq:g2QRM}). The red and green line corresponds to the scenario where the detectors coupled to the first (Eq.~\eqref{eq:dot_xa}) and second (Eq.~\eqref{eq:ddot_xa}) time derivatives of $\hat{x}_a$, respectively. All the calculations are done within the QRM for an intermediate pumping rate, $\Gamma/\gamma = 10^{-3}$.}
    \label{fig:g2_derivatives}
\end{figure}

Several recent contributions (Refs.~\cite{photon.blockade, garziano2017cavity}) have discussed alternative formulations of the intensity correlations, as defined by the time derivatives of $\hat{x}_a$:
\begin{equation}
    g^{(2)}(0)[\dot{\hat{x}}_a] = \frac{\braket{\dot{\hat{x}}_a^\dagger \dot{\hat{x}}_a^\dagger \dot{\hat{x}}_a \dot{\hat{x}}_a}_{ss}}{\braket{\dot{\hat{x}}_a^\dagger \dot{\hat{x}}_a}_{ss}^2},
\end{equation}
with
\begin{equation}
    \dot{\hat{x}}_a = \sum_{\mu, \nu; \omega_\nu > \omega_\mu} -i(\omega_\nu - \omega_\mu) \ket{\mu}\!{}_\text{R}~{}_\text{R}\!\bra{\mu} i(\hat{a}^\dagger - \hat{a}) \ket{\nu}\!{}_\text{R}~{}_\text{R}\!\bra{\nu},
    \label{eq:dot_xa}
\end{equation}
where $\ket{\mu}_\text{R}$ and $\ket{\nu}_\text{R}$ are the eigenstates of the QRM Hamiltonian (Eq. \eqref{eq:HQRM}) and $\omega_\nu > \omega_\mu$ their respective eigenvalues. This scenario where $g^{(2)}(0)$ depends on the $\dot{\hat{x}}_a$ operators describes an HBT interferometer with capacitive detectors that couple with the time derivative of the photons emitted by the cavity \cite{photon.blockade} (instead of coupling directly with the photons emitted in the cavity).

For completeness, we also consider the detectors of the HBT interferometer as coupled to the second-time derivative of the $\hat{x}_a$ operators (as studied in Ref.~\citenum{garziano2017cavity}),
\begin{equation}
    g^{(2)}(0)[\ddot{\hat{x}}_a] = \frac{\braket{\ddot{\hat{x}}_a^\dagger \ddot{\hat{x}}_a^\dagger \ddot{\hat{x}}_a \ddot{\hat{x}}_a}_{ss}}{\braket{\ddot{\hat{x}}_a^\dagger \ddot{\hat{x}}_a}_{ss}^2}.
\end{equation}
with 
\begin{equation}
    \ddot{\hat{x}}_a = \sum_{\mu, \nu; \omega_\nu > \omega_\mu} -(\omega_\nu - \omega_\mu)^2 \ket{\mu}\!{}_\text{R}~{}_\text{R}\!\bra{\mu} i(\hat{a}^\dagger - \hat{a}) \ket{\nu}\!{}_\text{R}~{}_\text{R}\!\bra{\nu}
    \label{eq:ddot_xa}
\end{equation}
Figure~\ref{fig:g2_derivatives} shows the evaluation of $g^{(2)}(0)[\dot{\hat{x}}_a]$ and $g^{(2)}(0)[\ddot{\hat{x}}_a]$, compared to $g^{(2)}(0)[{\hat{x}}_a]$ (the latter corresponds to the results in the main text, Eq.~\eqref{eq:g2QRM}). All formulations of $g^{(2)}(0)$ have  excellent agreement in the weak coupling regime ($\eta \lesssim 2.5 \times 10^{-2}$) and show the same qualitative trend for larger coupings (\textit{i.e.}, all formulations find a large bunching effect for the USC regime). However, some quantitative differences appear for $\eta \gtrsim 0.05$. These differences for large $\eta$ arise from the $(\omega_\nu - \omega_\mu)$ and $(\omega_\nu - \omega_\mu)^2$ factors in Eqs.~\eqref{eq:dot_xa} and~\eqref{eq:ddot_xa}, respectively. Intuitively, the influence of these extra factors is small in the weak coupling regime: In the WC regime ($\eta \lesssim 2.5 \times 10^{-2}$), the QRM eigenstates approach the description of the JCM polaritons, so that, $\ket{\mu}\!{}_\text{R}~{}_\text{R}\!\bra{\mu} i(\hat{a}^\dagger - \hat{a}) \ket{\nu}\!{}_\text{R}~{}_\text{R}\!\bra{\nu}$ in Eqs.~\eqref{eq:dot_xa} and~\eqref{eq:ddot_xa} is only non-zero for three conditions, either $\ket{\mu}_\text{R} = \ket{n-1\pm}_\text{R}$ with $\ket{\nu}_\text{R} = \ket{n\pm}_\text{R}$, $\ket{\mu}_\text{R} = \ket{n-1\mp}_\text{R}$ with $\ket{\nu}_\text{R} = \ket{n\pm}_\text{R}$ or $\ket{\mu}_\text{R} = \ket{0}_\text{R}$ with $\ket{\nu}_\text{R} = \ket{1\pm}_\text{R}$--- i.e., in the weak coupling, the $\dot{\hat{x}}_a$ operators describe only transitions between the nearest branch of eigenstates---. Thus, the prefactor $(\omega_\nu - \omega_\mu)$ can only include $(\omega_{n\pm} - \omega_{n-1\pm})$ or $(\omega_{n\pm} - \omega_{n-1\mp})$ terms. Furthermore, for small $\eta$, the eigenvalues of the two polaritons in each branch are very similar and differ by $(\omega_{n\pm} - \omega_{n-1\pm}) \approx (\omega_{n\pm} - \omega_{n-1\mp}) \approx \omega_0$, resulting in
\begin{equation}
    \lim_{\eta \to 0} g^{(2)}(0)[\dot{\hat{x}}_a] = \frac{\omega_0^4}{\omega_0^4}\frac{\braket{{\hat{x}}_a^\dagger {\hat{x}}_a^\dagger {\hat{x}}_a {\hat{x}}_a}_{ss}}{\braket{{\hat{x}}_a^\dagger {\hat{x}}_a}_{ss}^2} = g^{(2)}(0)[{\hat{x}}_a],
\end{equation}
or 
\begin{equation}
    \lim_{\eta \to 0} g^{(2)}(0)[\ddot{\hat{x}}_a] = \frac{\omega_0^8}{\omega_0^8}\frac{\braket{{\hat{x}}_a^\dagger {\hat{x}}_a^\dagger {\hat{x}}_a {\hat{x}}_a}_{ss}}{\braket{{\hat{x}}_a^\dagger {\hat{x}}_a}_{ss}^2} = g^{(2)}(0)[{\hat{x}}_a],
\end{equation}
As $\eta$ becomes larger, the eigenstates of the QRM differ from the JCM polaritons, and new transitions between eigenstates emerge. As a consequence, the $(\omega_\nu - \omega_\mu)$ factors can strongly vary from one transition to another, ultimately weighting each term differently in the series of Eqs.~\eqref{eq:dot_xa} and~\eqref{eq:ddot_xa}. Thus, for large $\eta$, $g^{(2)}(0)[{\hat{x}}_a]$, $g^{(2)}(0)[\dot{\hat{x}}_a]$, and $g^{(2)}(0)[\ddot{\hat{x}}_a]$ deviate.

\section{Derivation of the approximated expression for $g^{(2)}(0)$ in Eq.~\eqref{eq:g2_semi}}
\label{app:Derive_g2_semi}

Here we derive Eq.~\eqref{eq:g2_semi} in the main text. By considering the diagonal steady state approximation introduced in  Eq.~\eqref{eq:DiagonalSS} in the main text, we can write the expected value of any operator $\hat{O}$ in the steady state as:
\begin{align}
    \braket{\hat{O}}_{ss} &= \text{Tr}\{\hat{O} \hat{\rho}_{ss}\}  \\
    &\approx \text{Tr} \left\{\hat{O}\left(\sum_\nu R_\nu^\nu \ket{\nu}\!{}_\text{R}~{}_\text{R}\!\bra{\nu}\right)\right\} \nonumber \\
    &= \sum_\nu R_\nu~ {}_\text{R}\!\bra{\nu} \hat{O} \ket{\nu}\!{}_\text{R}. \nonumber
\end{align}
We then apply this formula to the expected value of $\braket{(\hat{x}_a^\dagger)^n (\hat{x}_a)^n}_{ss}$ ($n = 1$ and $n = 2$ for the numerator and denominator of $g^{(2)}(0)$, respectively), resulting in:
\begin{align}
    \braket{(\hat{x}_a^\dagger)^n (\hat{x}_a)^n}_{ss} &\approx \sum_\nu R_\nu~ {}_\text{R}\!\bra{\nu} (\hat{x}_a^\dagger)^n (\hat{x}_a)^n \ket{\nu}\!{}_\text{R} \nonumber \\
    &=\sum_\nu R_\nu~ {}_\text{R}\!\bra{\nu} (\hat{x}_a^\dagger)^n \hat{\ident}(\hat{x}_a)^n \ket{\nu}\!{}_\text{R},
\end{align}
where we have included in the last step the $\hat{\ident}$ identity matrix. Because the eigenstates of the QRM Hamiltonian are orthonormal we can write the identity matrix as $\hat{\ident} = \sum_\mu \ket{\mu}\!{}_\text{R}~{}_\text{R}\!\bra{\mu}$, and thus:
\begin{align}
    &\sum_\nu R_\nu~ {}_\text{R}\!\bra{\nu} (\hat{x}_a^\dagger)^n \hat{\ident}(\hat{x}_a)^n \ket{\nu}\!{}_\text{R} \nonumber\\
    &=\sum_{\mu, \nu} R_\nu~ {}_\text{R}\!\bra{\nu} (\hat{x}_a^\dagger)^n \ket{\mu}\!{}_\text{R}~{}_\text{R}\!\bra{\mu}(\hat{x}_a)^n \ket{\nu}\!{}_\text{R} \nonumber \\
    &=\sum_{\mu, \nu} R_\nu |{}_\text{R}\!\bra{\mu} (\hat{x}_a)^n \ket{\nu}_\text{R}|^2,
    \label{eq:derivation_last}
\end{align}
where we have used the property $\braket{b|\hat{O}^\dagger|a} = (\braket{a|\hat{O}|b})^*$. Applying Eq.~\eqref{eq:derivation_last} for $n = 1$ (denominator of $g^{(2)}(0)$) and $n = 2$ (denominator of $g^{(2)}(0)$) directly results in Eq.~\eqref{eq:g2_semi}.

\section{Effects of the Ohmic bath spectrum}

\begin{figure}[t!]
    \centering
    \includegraphics[width=\linewidth]{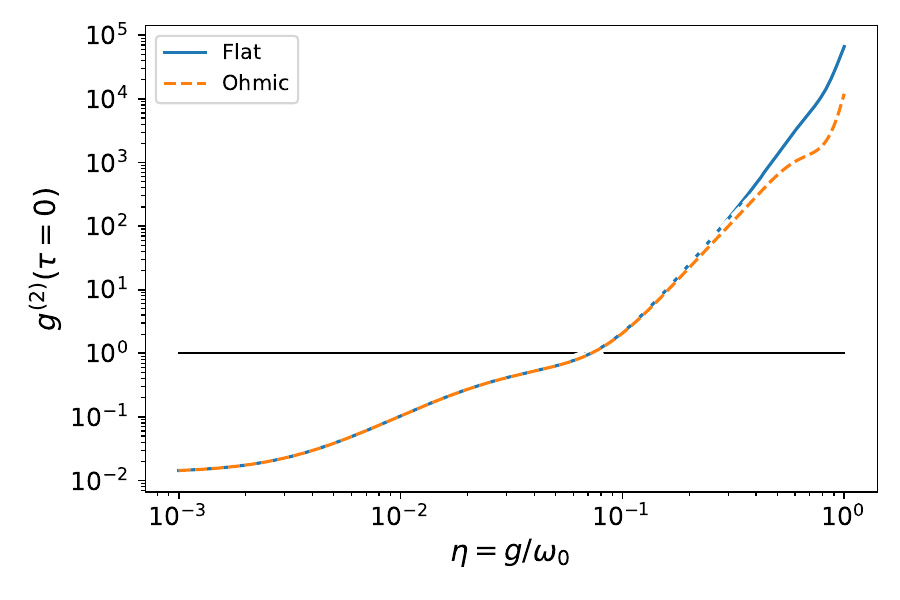}
    \caption{Intensity correlations calculated using the QRM, assuming a bath with a flat spectra response (solid blue lines), and an Ohmic bath (orange dashed lines). In the latter model we replace the constant decay rates for the TLS and the cavity ($\gamma$ and $\kappa$) with a rate that is dependent on the transition energy (see Eq.~\eqref{eq:bath}). The remaining parameters are as in Fig.~\ref{fig:setup_problem}(b).}
    \label{fig:Ohmic}
\end{figure}

Few models of the bath have been discussed in the literature, including the spectrally flat (Fig. 5 in Ref.~\citenum{Hughes2022gauge}), or Ohmic bath (where the density of states is proportional to frequency, see e.g.~\cite{settineri2018dissipation}). These publications identified quantitative differences to the emission spectra between the two bath models deep in the USC regime, where the emission spans across a wide spectral range. 

In Fig.~\ref{fig:Ohmic}, we compare the intensity correlations calculated using the flat and Ohmic bath models. To this end, we revisit the setup discussed in Fig.~\ref{fig:setup_problem}(b), and calculate $g^{(2)}(0)$ using the flat bath (where the decay rates $\gamma$ and $\kappa$ are constant for each transition frequency $\omega_{kl}$; blue solid lines), and the Ohmic bath (where we replace 
\begin{equation}
    \gamma \rightarrow \gamma \frac{\omega_{kl}}{\omega_0},\quad \kappa \rightarrow \kappa\frac{\omega_{kl}}{\omega_0},\label{eq:bath}
\end{equation}
and plot the results with orange dashed lines). 

The differences between the two models are negligible for the weakly coupled system, and both models predict the strong bunching in the USC. Quantitative changes can be identified in the USC, consistent with the studies of the spectra in QRM~\cite{settineri2018dissipation,Hughes2022gauge}. These changes are mostly due to the corrections to the TLS decay rates.

We have also verified numerically that the same agreement between the two bath models, with deviations occurring in the USC, can be observed for the entire range of pumping rates discussed in this work (not shown here).

\section{Experimental feasibility of bunching measurments}\label{app:experimental}

In the limit of weak coupling and high cooperativity, the rate of photon emission from the cavity can be estimated in several ways: 
\begin{enumerate}
    \item in the steady-state the system balances pumping and emission; therefore, neglecting non-radiative processes, we can expect the photon emission rate as approximately equal to the incoherent pumping rate $\Gamma$,
    \item the bulk of the emission should come from the lowest-order, and highest-population polaritonic states with appreciable cavity components, like the $\ket{1+}_\text{R}$ or $\ket{1-}_\text{R}$; therefore, the photon emission rate is approximately given by the product of this population, and the emission rate from that state.
\end{enumerate}
For the parameters where we expect a deviation between the predictions of the JCM and QRM (see Fig.~\ref{fig:g2_colormap}), for example $\eta = 0.1$ or $C = 800$, and $\Gamma/\gamma = 10^{-3}$, the emission rates can be estimated from these two methods as $\Gamma = 10^{-6} \omega_0 \sim 2\pi\times 2~10^8~\text{s}^{-1}$, and $\kappa R_{1+} \sim (2\pi\times 10^{13}~\text{s}^{-1})\times 10^{-5} = 2\pi\times 10^8~\text{s}^{-1}$ (see Fig.~\ref{fig:Populations} for estimates of $R_{1+}$). With the perfect collection efficiency, we would therefore expect emission every 1~ns.

Bunching predicted by the QRM, and measured as $g^{(2)}>1$, should vanish on the timescales dictated by the decoherence of the emitter – here given by the product of cooperativity and the emitter decay rate $C\gamma \approx 800 ~ (2\pi \times 2~ 10^{11}~\text{s}^{-1}) \approx 2\pi\times 2~10^{14}~\text{s}^{-1}$, that is the bunching would vanish within approximately 0.1~ps. The probability of a coincidence detection within that 0.01~ps window is of the order of 0.01~ps/1~ns $\approx 10^{-5}$, which yields the coincidence rate of $10^{-5} \times (2\pi~ 10^{8}~\text{s}^{-1}) = 2\pi \times 10^3~\text{s}^{-1}$.

This coincidence rate can be also estimated by assuming that the two-photon emission is exclusively due to the relaxation from the $\ket{3-}_\text{R}$ state (see discussion in Section~\ref{sec:origin.of.bunching} and Eqs.~\eqref{eq:DiagonalSS} and \eqref{eq:g2_semi}), and given by a product of its population ($\sim 10^{-10}$, see Fig.~\ref{fig:Populations}) and emission rate taken as $C\gamma \approx 2\pi \times  2~10^{14}~\text{s}^{-1}$. This approach yields a similar estimate of coincidence rate of $2\pi \times 2~10^{4}~\text{s}^{-1}$.

While this coincidence rate was estimated with generous assumptions about the collection efficiency, it remains several orders magnitude larger than the rates reported in contributions on the characterisation of statistics of faint emission from nonclassical emitters~\cite{Kasperczyk:15,galland}. 

\end{document}